\label{packages}

\documentclass[10pt,journal,compsoc]{IEEEtran}

\ifCLASSOPTIONcompsoc
  \usepackage[nocompress]{cite}
\else
  \usepackage{cite}
\fi

\usepackage{enumitem}
\usepackage{textcomp}
\usepackage{amsmath}
\usepackage{subfig}
\usepackage{graphicx}
\usepackage{color, soul}
\usepackage[linesnumbered,ruled]{algorithm2e}
\usepackage{algpseudocode}
\usepackage{array}

\algnewcommand\algorithmicforeach{\textbf{for each}}
\algdef{S}[FOR]{ForEach}[1]{\algorithmicforeach\ #1\ \algorithmicdo}
\newcolumntype{P}[1]{>{\centering\arraybackslash}p{#1}}

\begin{document}
\title{A Scheduling Algorithm to Maximize Storm Throughput in Heterogeneous Cluster}

\author{Hamid~Nasiri,~
        Saeed~Nasehi,~
        Arman~Divband,~
        and~Maziar~Goudarzi,~\IEEEmembership{Member,~IEEE}
\IEEEcompsocitemizethanks{
	\IEEEcompsocthanksitem H. Nasiri, S. Nasehi, A. Divband and M. Goudarzi are with the Department of Computer Engineering, Sharif University of Technology, Tehran, Iran.\protect\\
E-mail:\{hnasiri,saeednasehi,divband\}@ce.sharif.edu, goudarzi@sharif.edu
}
}

\markboth{arXive, 2020}
{Nasiri \MakeLowercase{\textit{et al.}}: A Scheduling Algorithm to Maximize Storm Throughput in Heterogeneous Cluster}

\IEEEtitleabstractindextext{%
\begin{abstract}
In the most popular distributed stream processing frameworks (DSPFs), programs are modeled as a directed acyclic graph. This model allows a DSPF to benefit from the parallelism power of distributed clusters. However, choosing the proper number of vertices for each operator and finding an appropriate mapping between these vertices and processing resources have a determinative effect on overall throughput and resource utilization; while the simplicity of current DSPFs' schedulers leads these frameworks to perform poorly on large-scale clusters.
In this paper, we present the design and implementation of a heterogeneity-aware scheduling algorithm that finds the proper number of the vertices of an application graph and maps them to the most suitable cluster node. We start to scale up the application graph over a given cluster gradually, by increasing the topology input rate and taking new instances from bottlenecked vertices. Our experimental results on Storm Micro-Benchmark show that 1) the prediction model estimate CPU utilization with 92\% accuracy. 2) Compared to default scheduler of  Storm, our scheduler provides 7\% to 44\% throughput enhancement. 3) The proposed method can find the solution within 4\% (worst case) of the optimal scheduler which obtains the best scheduling scenario using an exhaustive search on problem design space.
\end{abstract}

}

\maketitle

\IEEEdisplaynontitleabstractindextext
\IEEEpeerreviewmaketitle

\IEEEraisesectionheading{\section{Introduction}\label{sec:introduction}}

\IEEEPARstart{W}{ith} the growing of stream processing applications such as network monitoring, online machine learning, signal processing, and sensor-based monitoring, the data generation model is potentially an infinite sequence of data. Because of the huge amount of data and high speed of arrival, this unlimited sequence of data needs to be processed uninterruptedly. Apache Storm [1] is among the most popular distributed stream processing frameworks for such applications. In Storm, applications are modeled as a directed acyclic graph (DAG), named \textit{topology}. To run a topology in Storm, the user determines the structure of execution topology graph in which the number of instances (tasks) for each component of the application is specified. Then the Storm's scheduler maps all tasks of the topology to processing elements (\textit{PE}s). For task assignment, Storm uses a \mbox{Round-Robin} algorithm as its default scheduler, and hence it maps tasks to the existing PEs, regardless of their computing power [2].
\par
In stream processing, depending on the computation requirements and the input rate of each application we must take sufficient set of processing resources, to prevent CPU overloading. There are many streaming applications in which incoming data arrives with a constant rate, such as surveillance applications, weather analysis, predictive maintenance and any other system that its sensors collect data periodically. When the input rate of a streaming system has a constant value, its required processing power remains steady, therefore we can optimize its mapping to the target cluster in order to efficiently use processing resources. Especially, when a streaming application is running on a heterogeneous cluster which has different types of processors, proper selection of the number of instances for each component and assignment of these instances to PEs are too important; because these both affect overall throughput and resource utilization. Application to architecture mapping is an old classic design automation problem, but there is a number of features that make Storm optimization unique and new among them: (i) the number of instances of each component is determined by the user and can be set based on the executing cluster; (ii) computation load of each instance is not constant, and can be tuned by changing the rate of dispatching input tuples to it; (iii) input rate of each instance determines workload of its downstream instance(s). Thus, new techniques that take these features into account can produce higher throughput implementations.
\par
Several scheduling methods such as [3] ,[4], [5], [6] are proposed to improve the naive Storm scheduler, but most of them focus on resource management in homogeneous clusters. None of these works effectively considers heterogeneity of cluster nodes, and they perform poorly on heterogeneous clusters, which are the prevalent case in real world data centers. Other researches such as [7], [8], [9], [10] present some scaling methods to dynamically scale up or down a Storm cluster depending on processing demand or input data rate. In both scheduling and scaling solutions, a primary user graph is submitted to Storm to be executed and the number of instances for each component of this graph is determined by user.
\par
\begin{figure}[!t]
	\centering
	\includegraphics[width=3.5in]{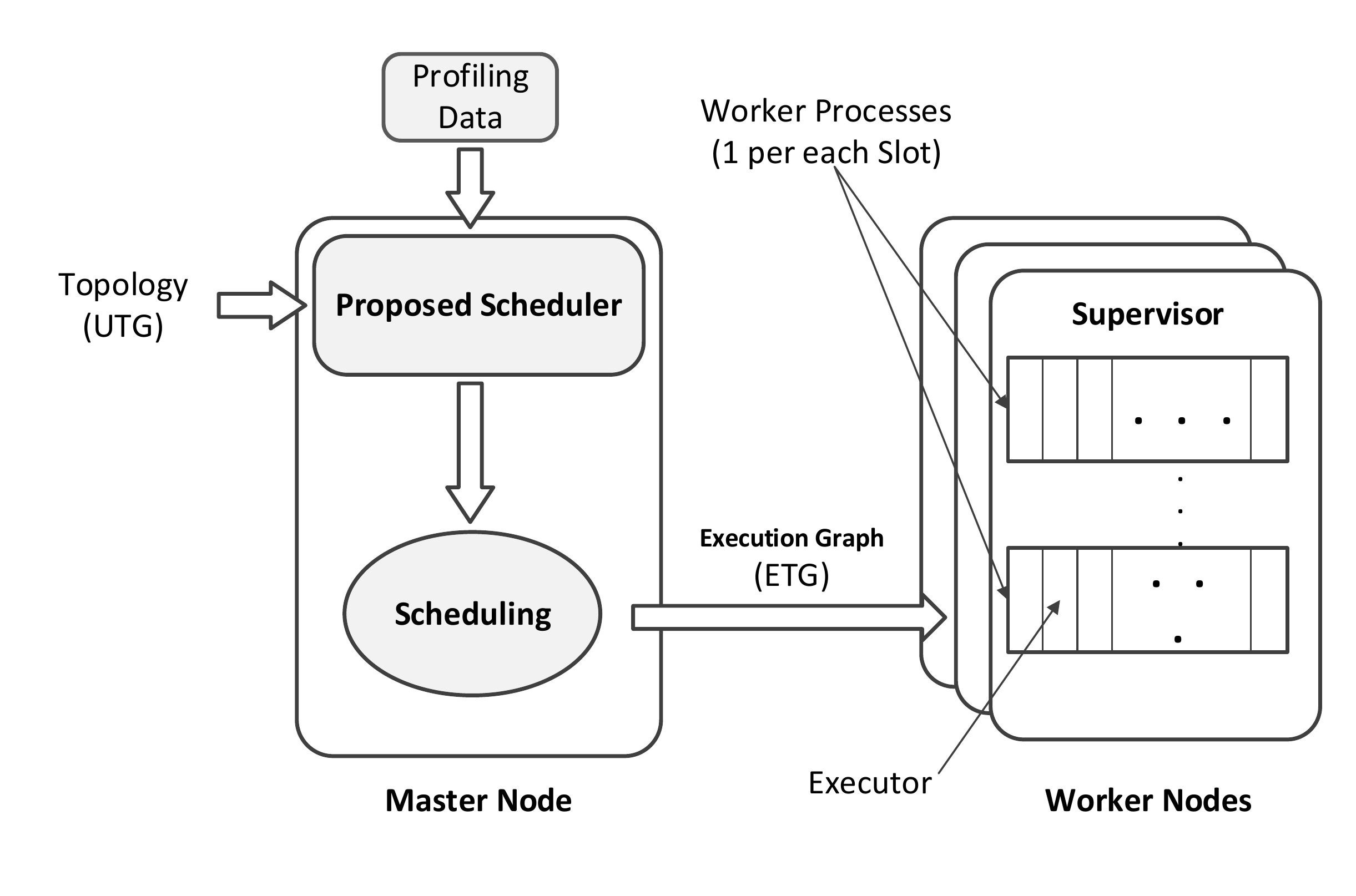}
	\caption{Architecture overview of Apache Storm with proposed scheduler}
	\label{big_picture}
\end{figure}

In this paper, we propose a novel scheduling algorithm, which 1) calculates the near-optimum number of instances for each component to create the fitting topology for a specific heterogeneous cluster; and 2) chooses the most appropriate machine for each task, by considering its processing characteristics. For a given heterogeneous cluster, the proposed algorithm tries to fully utilize the processing resources while it prevents over-utilization of them. To guarantee that no over-utilization happens, we present a heterogeneity-aware model to predict CPU utilization of each task, running on a specific machine. Fig. \ref{big_picture} shows the architecture overview of Apache Storm with our heterogeneous scheduler on a cluster of $n$ worker node. We modified the source code of Storm and implemented a new heterogeneity-aware scheduler. Our scheduler takes user topology graph and some profiling data as inputs and results in an execution topology graph. Then it maps the output graph to the worker nodes with regards to their capability.
\par
To evaluate the efficiency of the proposed algorithm, different types of topologies from Micro-Benchmark [6] and Storm-Benchmark [15] are executed on several clusters of heterogeneous machines. As there is no similar scheduling algorithm, which produces the execution graph according to the computing power of heterogeneous machines, we compare our work with the optimum scheduler. The optimum scheduler obtains the best execution graph using an exhaustive search on the problem design space. The experimental results show that our algorithm provides 7\% to 44\% throughput enhancement in comparison with default scheduler of Storm. Moreover, real occupied CPU utilizations show the proposed method utilizes all processing resources efficiently and comparison of them with the predicted values demonstrates that our prediction model can estimate CPU utilization with over 92\% accuracy.
\par
This article makes the following major contributions:
\begin{enumerate}[label=\arabic*.]
	\item To the best of our knowledge, it provides the first scheduling algorithm for Apache Storm in which a fitting topology graph for a specific heterogeneous cluster is created while resource heterogeneity is taken into account.
	\item Our proposed algorithm leads general-purpose stream processing applications to reach the maximum achievable throughput of a heterogeneous cluster.
	\item We provide the implementation of the proposed algorithm in Apache Storm and evaluate it using topologies from Micro-Benchmark[6] and Storm-Benchmark[15] for both throughput and resource usage efficiency.
\end{enumerate}	
\par
The remainder of this paper is organized as follows. Section 2 explains some primary information about stream processing and Apache Storm architecture. Section 3 shows our motivation for considering resource heterogeneity. Section 4 clarifies problem definition. Section 5 describes the details of the proposed algorithm. Section 6 reports the evaluation of our prediction model and scheduling algorithm. Section 7 reviews related works. Finally, Section 8 discusses conclusion and future works.

\section{Background}

\subsection{Stream Processing}
Big data processing can be divided into two main categories: Batch processing and Stream processing. In Batch or off-line processing, data is prepared before processing while in the stream or online processing, a possibly infinite sequence of data items, generated continuously in time and must be processed in real-time. Apache Hadoop is one of the most popular frameworks for batch processing, which uses \textit{MapReduce} programming model. There are several large-scale computing architectures customized for batch processing of big data applications [11]; however, they are not suitable for stream data processing because, in the MapReduce paradigm, all input data needs to be stored on a distributed file system like HDFS, before start to process. To address large-scale stream processing problems, many distributed frameworks have emerged; however, Apache Storm is the best-known stream processing framework for low latency and high throughput applications, among them all [22].

\subsection{Apache Storm Programming Model}
\begin{figure*}[!t]
	\centering
	\subfloat[A sample user graph with 1 spout and 4 bolts]{\includegraphics[width=2.2in]{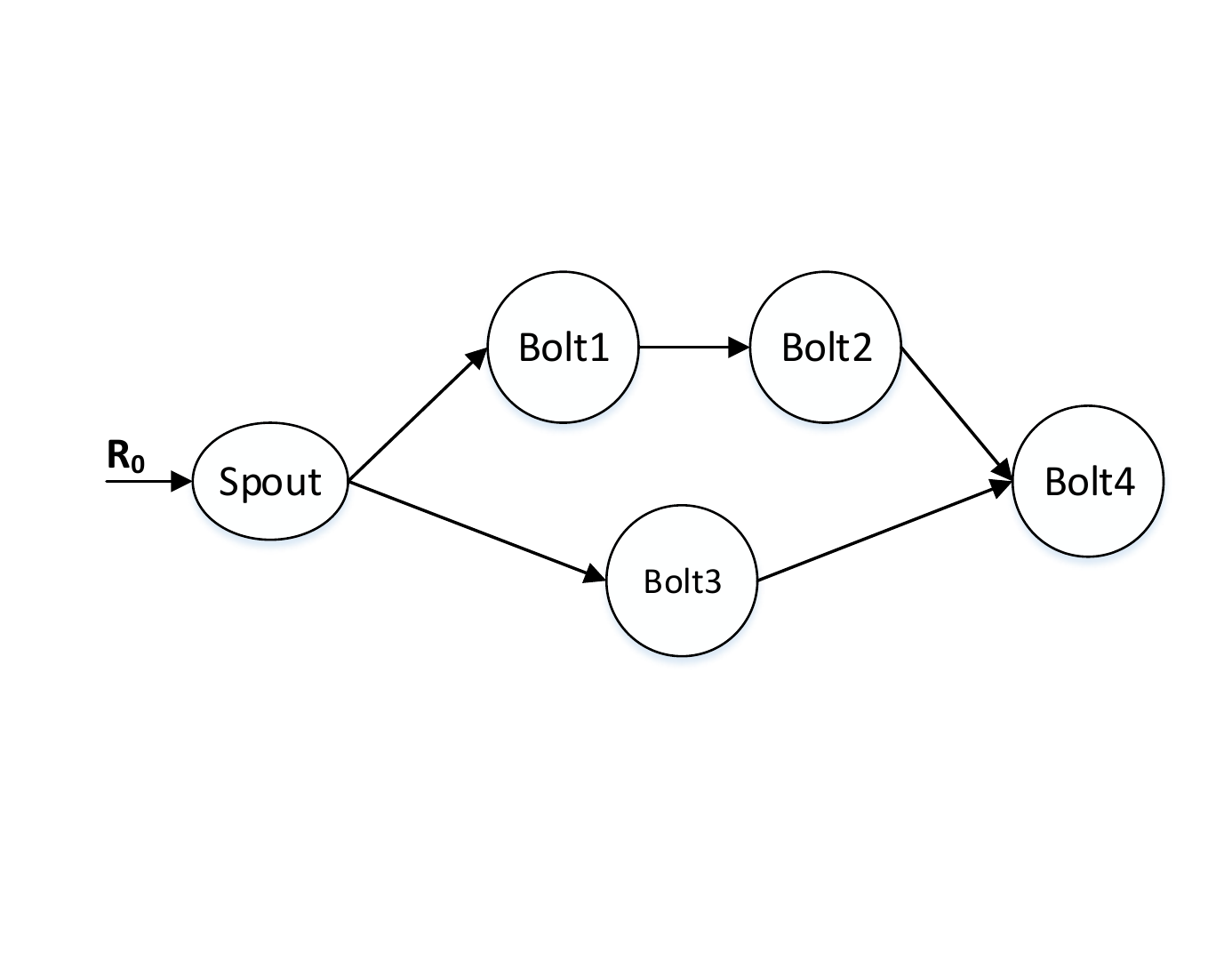}
		\label{user_topology}}
	\hfil
	\subfloat[An execution graph for Fig. \ref{user_topology} with different number of instances for each bolt]{\includegraphics[width=3in]{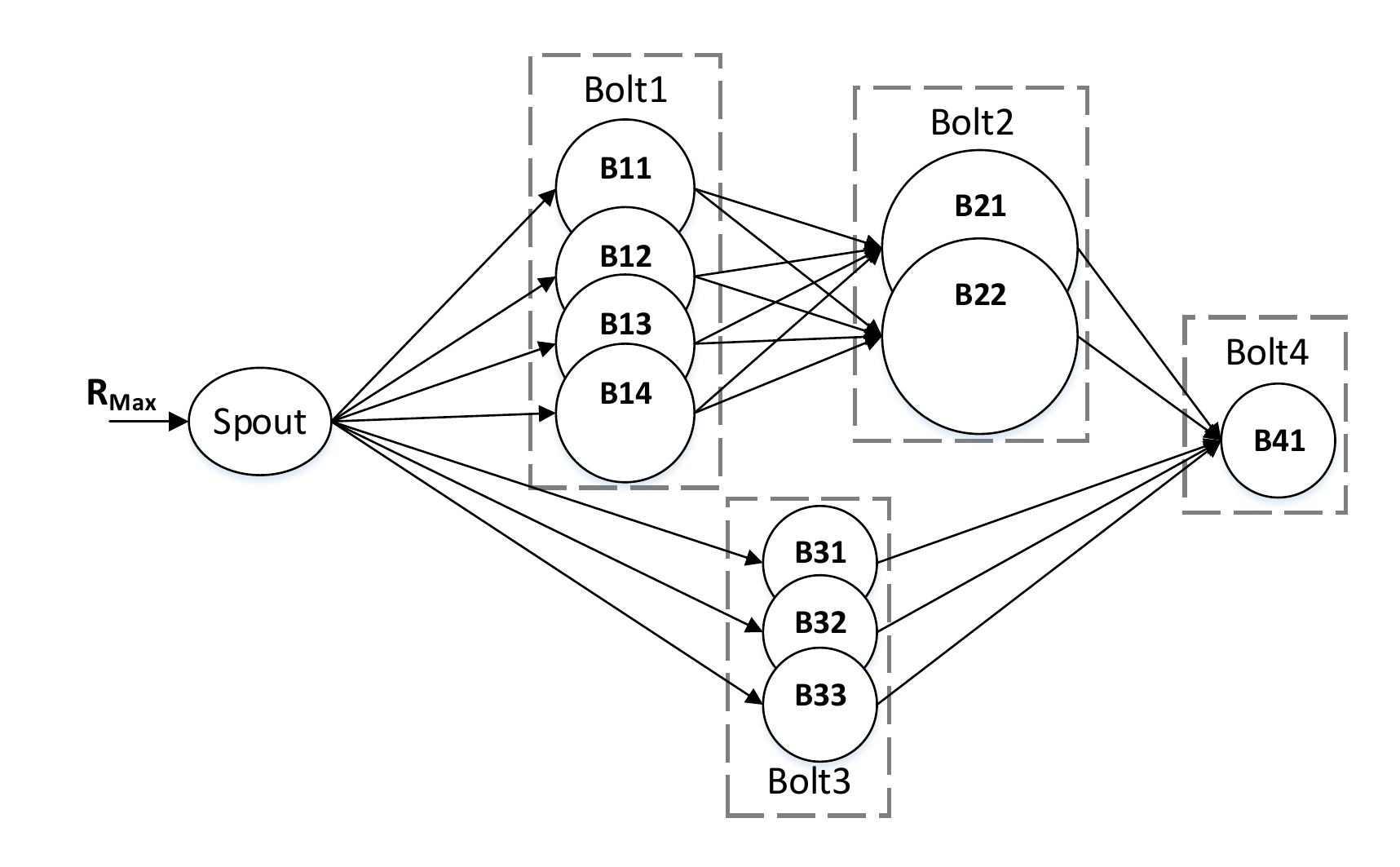}
		\label{exec_topology}}
	\hfil
	\subfloat[Task assignment of Storm's default scheduler for execution graph Fig. \ref{exec_topology}]{\includegraphics[width=1.7in]{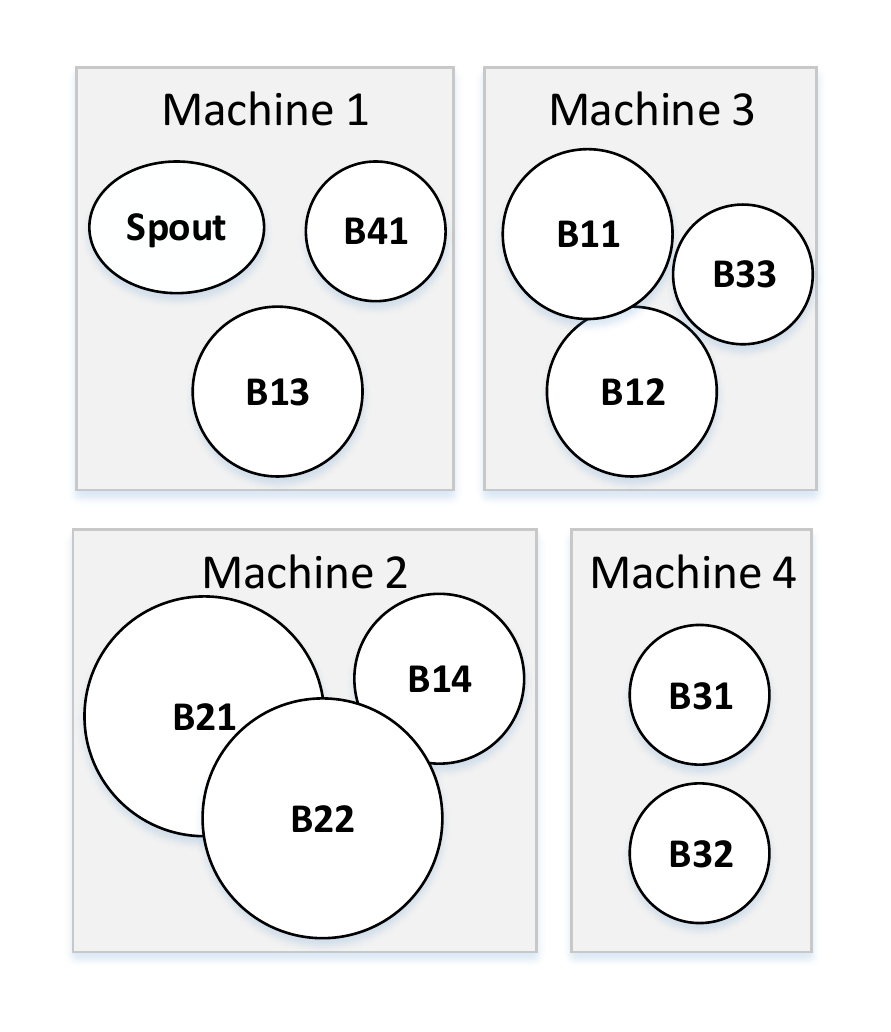}
		\label{running_topology}}
	\caption{User graph (a), execution graph (b), and task  assignment (c) of a sample topology}
	\label{topology}
\end{figure*}

Storm is an open source, real-time and distributed processing platform that processes unbounded stream of data. In Storm, each program is modeled as a directed acyclic graph (DAG) called \textit{topology}. A topology consists of two types of components. The component responsible for input stream production is \textit{spout} and the component in charge of data processing is \textit{bolt}. This primary topology is known as user topology graph (\textit{UTG})[12]. Fig. \ref{user_topology} depicts a user graph with one spout and four bolts.
\par
Each component of the user topology graph may have several instances (\textit{tasks}) that determines its parallelism degree. All instances of a component execute the same operation on different parts of the upstream data. A topology with distinct parallelism degrees, forms the execution topology graph (\textit{ETG}) [12]. Fig. \ref{exec_topology} illustrates an example execution graph corresponding to the user graph in Fig. \ref{user_topology}. In this figure, the parallelism degrees of Bolt1, Bolt2, Bolt3, and Bolt4 are 4, 2, 3 and 1 respectively. We assumed each bolt of the topology has variable size because each bolt occupies different CPU utilization, depending on its input rate and computation requirements. Structure of the execution topology graph dramatically affects overall throughput and average tuple processing time, which are two major considerations in stream processing. 
\par
Once a topology is assigned to processing elements (Fig. \ref{running_topology} shows a sample assignment of execution topology graph Fig. \ref{exec_topology}), the spout(s) brings data into the system and sends them to the bolt(s). As processing progresses, one or more bolts may write data out to a database or file system, send a message to another bolt (s), or otherwise make the results of the computation available to the user [12]. 

\subsection{Apache Storm Execution Architecture}
A Storm cluster consists of two kind of nodes: \textit{master} node and \textit{worker} nodes. Master node runs a daemon called \textit{Nimbus}, which is the central component of Apache Storm. The main responsibility of nimbus is distributing tasks on the worker nodes. Worker nodes do the actual execution of the topology [13]. Each worker node (machine) runs a daemon called \textit{Supervisor}, which executes a portion of the topology. The configuration of worker node determines how many \textit{worker processes} (correspond to \textit{slots}) it provides for the cluster. Each worker process runs a Java Virtual Machine (JVM), in which several threads (known as \textit{executor}) are running on.  
\par 
A scheduling solution specifies how to assign tasks to worker nodes. Apache Storm has a default scheduler, which uses a simple Round-Robin method to map executors to worker processes. Then, it evenly distributes worker processes to available slots on worker nodes. Fig. \ref{running_topology} shows how Storm default scheduler distributes all tasks of the execution topology graph of Fig. \ref{exec_topology} among four different machines of a heterogeneous cluster. 
\par

\section{Motivation}
In the last decade, CPUs are improved in speed and computing power, mainly by increasing in frequency of their clocks or developing new architectures. When a data center is scaled up over time, it would include several generations of processors and it makes the data center to have a heterogeneous processing environment. On the other hand, big data applications are mostly processed on large-scale computer systems such as data centers nowadays. Therefore, in practice, we have heterogeneous clusters as infrastructures of big data processing. 
\par
\begin{figure*}[!t]
	\centering
	\subfloat[Linear topology]{\includegraphics[width=2.3in]{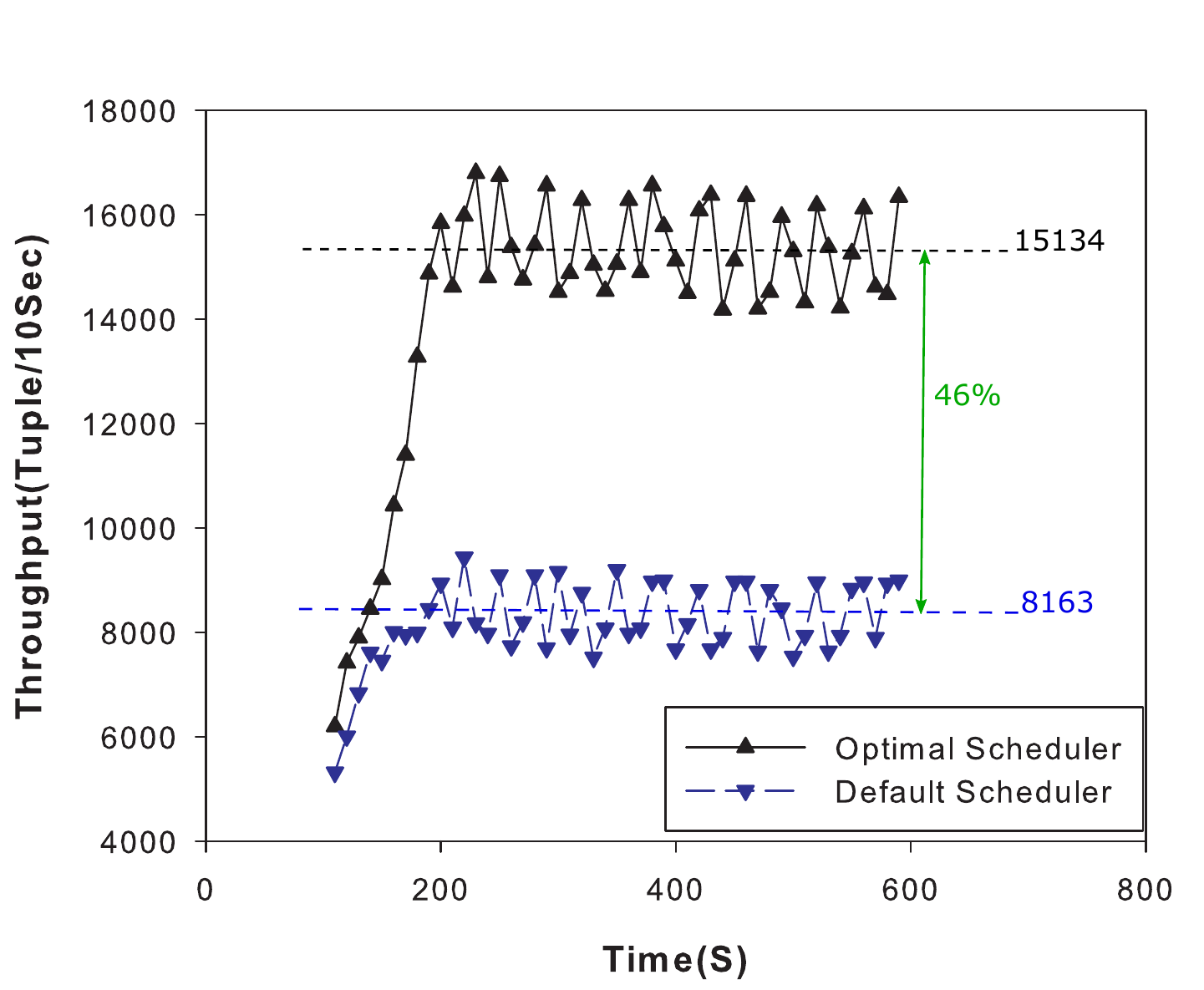}
		\label{motiv_linear}}
	\hfil
	\subfloat[Diamond topology]{\includegraphics[width=2.3in]{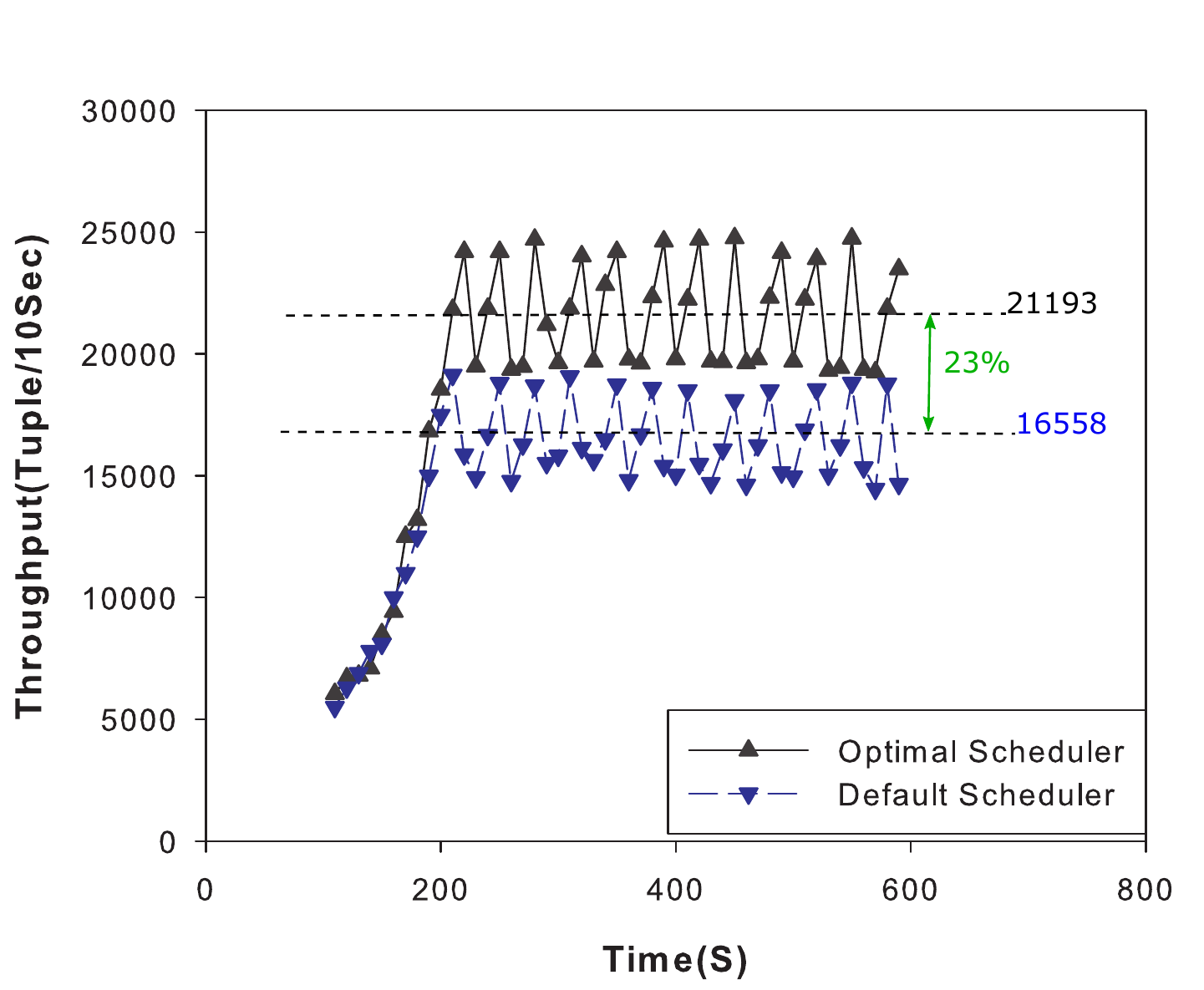}
		\label{motiv_diamond}}
	\hfil
	\subfloat[Star topology]{\includegraphics[width=2.3in]{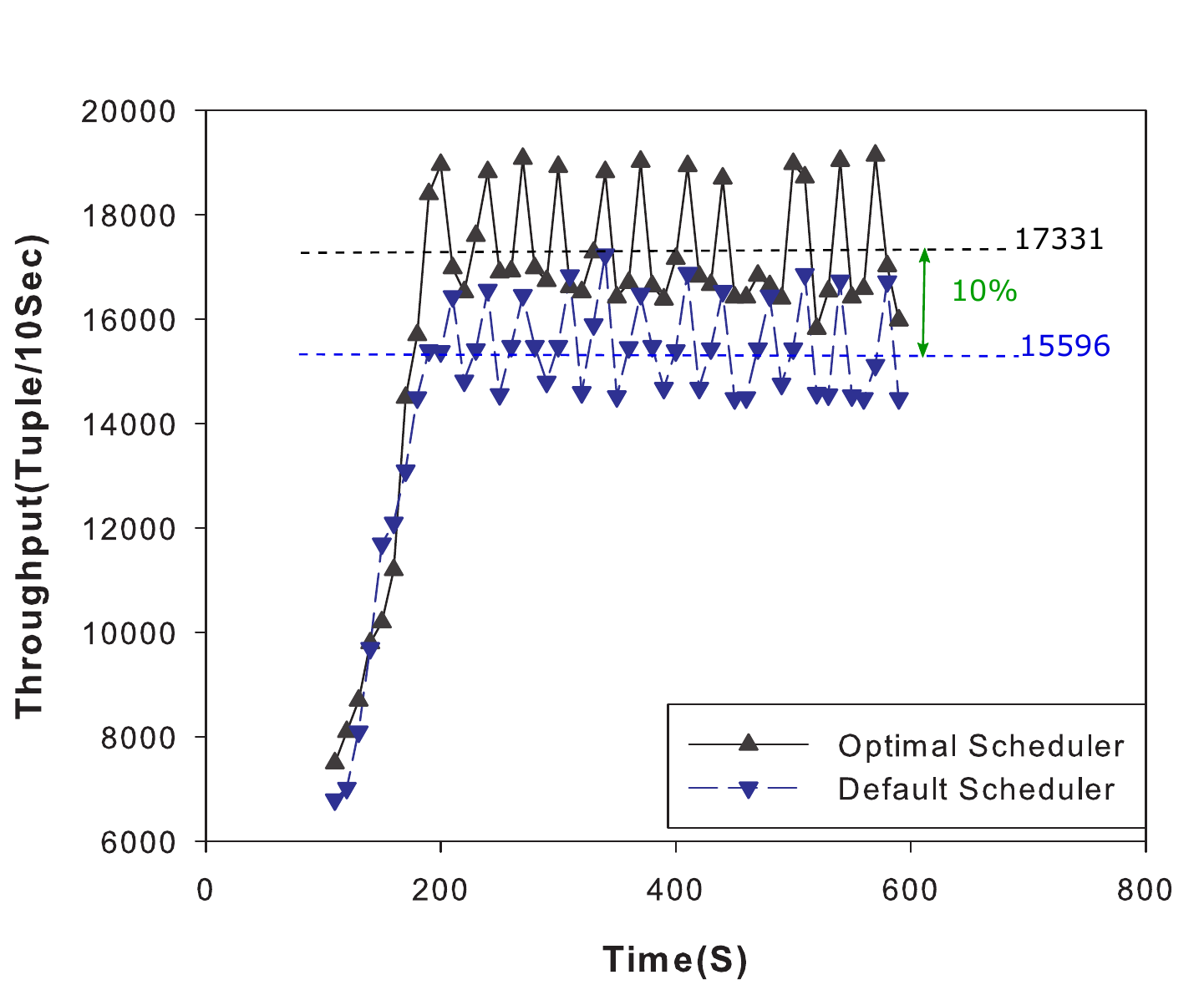}
		\label{motiv_star}}
	\caption{Throughput comparison of default and optimal schedulers for different topologies}
	\label{motiv_results}
\end{figure*}

Despite the pervasiveness of heterogeneous clusters in the existing data centers, Apache Storm does not consider resource heterogeneity in its scheduling solutions. Therefore, its schedulers perform same task assignment in both homogeneous and heterogeneous systems. To show the inefficiency of Storm default scheduler in heterogeneous environments, we ran three basic topologies from Storm Micro-Benchmark [6] on a cluster of three different machines (the experimental setup is described in Section 6). Fig. \ref{motiv_results} represents the total throughput of running intended typologies when their tasks are assigned to the cluster nodes by the Storm default scheduler and optimal scheduler.
\par
The optimal scheduler is a brute-force algorithm, which performs an exhaustive search on the task-assignment design space to find the best placement for all tasks of the topology on the machines of the target cluster. This algorithm calculates the overall throughput for all possible placements and then chooses the placement, which results in maximum overall throughput for the topology. As we can see in Fig. \ref{motiv_results}, scheduling scenario has a significant impact on the overall throughput, and there is a remarkable gap between the current and the most achievable throughput. Therefore, at first sight, we may prefer to use optimal scheduler instead of Storm default scheduler to get higher throughput and efficiency, but clearly, the optimal scheduler has much higher execution time. 
\par
To determine the time complexity of the optimal scheduling algorithm, we have to limit the total number of topology tasks. Assuming each machine can execute at most $k_j$ tasks simultaneously, the total number of tasks is limited to the processing power of machines and hence, all possible non-negative integer solutions for task assignment can be obtained using the following equation:
\begin{IEEEeqnarray}{c}
x_1 + x_2 + x_3 + ... + x_n \leq \sum_{j=1}^{m} k_j \\
s.t. \hspace{0.4cm}\forall i = 1 ... n \hspace{0.4cm} x_i \geq 1 \IEEEnonumber
\end{IEEEeqnarray}
\par	
In this equation, $m$ is the number of candidate machines for task placement, $k_j$ is the maximum number of tasks which can be executed on $j^{th}$ machine simultaneously, $n$ is the total number of components and $x_i$ is the total number of tasks for $i^{th}$  component. By solving this equation, the time complexity of optimal scheduling algorithm is obtained $ O(c(\sum_{j=1}^{m} k_j, n))$. For example, running the optimal scheduler for mapping a topology with four bolts $(n=4)$ on three machines $(m=3)$ with same processing power $(\forall j = 1, 2, 3\hspace{0.3cm} k_j = 10)$ checks \textit{27405} possibilities and lasts about \textit{18} hours, by a server with four Xeon 5560 processors and 8 GB of memory. When we have a heterogeneous cluster in which the machines have different processing powers, this time is increased exponentially.
\par 
So, to address the scheduling problem in heterogeneous systems, we need a new scheduler that performs faster task-assignment than the optimal scheduler and results in higher throughput than the default scheduler does. In the following section, we discuss problem definition and its unique considerations then we explain our scheduling method in detail, in section 5.
	
\section{Problem definition}
The gist of the problem we are trying to solve is to find an adequate number of tasks and how to best assign them to machines in order to maximize the overall throughput. Each component has a certain processing requirement that determines how many instances must be taken from it. On the other hand, in a cluster of heterogeneous machines, the number of instances for each component significantly depends on the type of the machine, which the instances are assigned to. Given these processing requirements and the processing power of machines, how can we create enough tasks and schedule them such that fully utilize the cluster and hence maximize overall throughput; while no machine is exceeding its processing capacity. The parameters used in our approach are summarized in Table \ref{symbols}.
\begin{table}[!t]
	\renewcommand{\arraystretch}{1.5}
	\caption{Terms and definitions of the approach}
	\label{symbols}
	\centering
	\begin{tabular}{c|p{2in}}
		\bfseries Symbols & \bfseries Descriptions\\
		\hline\hline
		$N_{T_i}$ & Number of available machines with type $T_i$\\
		$N_{C_i}$ & Total number of instances of component $i$\\
		$m$ & Total number of worker nodes (machines)\\
		$n$ & Number of topology's components\\
		$PT_{iw}$ & Processing throughput of task $i$ running on machine $w$\\
		$MAC_w$ & Available CPU capacity of machine $w$\\
		$TCU_{ij}$ & Occupied CPU utilization by $i^{th}$ task on machine $j$\\ 
		$e_{ij}$ & Average tuple execution time of task $i$ on machine $j$ \\
		$MET_{ij}$ & Miscellaneous execution time of Storm for task $i$ running on machine $j$\\
		$R_0$ & Topology initial input rate\\
		$\alpha_i$ & Tuple division ratio of $i^{th}$ component\\
		$IR_i$ & Input rate of $i^{th}$ task\\
		$PR_i$ & Processing rate of $i^{th}$ task\\
		$OR_i$ & Output rate of $i^{th}$ task\\
	\end{tabular}
\end{table}

\subsection{Problem Statement} 
There is a finite number of heterogeneous machines on the target cluster. In general, the number of available machines is specified by the user. In the context of our system, each machine (worker node) has one worker process, which has a particular processor architecture and a specific processing budget. Furthermore, we assume each task from the execution topology graph is mapped to one executor, in other words, the total number of tasks is equal to the total number of executors. 
\par
Let $ TC = (T_1, T_2, T_3, ...)$ be the set of all types of machines in the target cluster, and $N_{T_{i}}$ shows the number of available machines with type $T_i$ , on that cluster. For a heterogeneous cluster with $T$ types of machines, the summation of $N_{T_{i}}$s $(i = 1 ... T)$ shows the total number of machines ($m$). Also let $ UTG = (c_1, c_2, c_3, ...)$ be the set of all components within a user topology graph, and similarly, $N_{C_{i}}$ says how many instances each component has. Here the summation of $N_{C_{i}}$s shows the total number of tasks ($n$).
	
\subsection{Problem Formulation}
The objective of our problem is to maximize overall throughput by finding the near-optimal number of instances for each component and appropriate placement of these instances on a cluster of heterogeneous machines while at the same time not making a scheduling that will over-utilize the CPUs of machines. In Storm, the overall throughput of a topology is assumed as the summation of processing throughput of all tasks [6]. Since each component may have several instances, for each component we sum the processing throughput of its instances $(PT_{iw}$s$)$ and then these values are added together to obtain the overall throughput of a topology. To make sure that no machine is over-utilized, we put a constraint on \textit{Machine Available CPU capacity} $(MAC_w)$. For each machine, this constraint checks that the summation of CPU utilization of its tasks $(TCU_s)$ not to exceed the total utilization budget of its CPU. The value of $(MAC)$s is assumed to be 100 at the beginning. Another constraint checks if all components have at least one instance, to create the minimal execution topology graph. In other words, the objective of our problem is defined as follow: 
\begin{IEEEeqnarray}{rCl}
	maximize\hspace{0.4cm} &&\sum_{j=1}^{n} \sum_{k=1}^{N_{C_{j}}} \sum_{w=1}^{m} PT_{iw} \\
	subject\hspace{0.1cm} to \hspace{0.4cm} &&\forall w=1 ...m  \hspace{0.6cm} MAC_w \geq 0 \nonumber\\
	&&\forall j=1 ... n  \hspace{0.8cm} N_{C_{j}} \geq 1 \nonumber
	\label{objective}
\end{IEEEeqnarray}
where
\begin{IEEEeqnarray}{c}
	i = (\sum_{l=1}^{j} N_{C_{l}}) + k  
 	\label{index}
\end{IEEEeqnarray} 
\begin{IEEEeqnarray}{c}
PT_{iw} = \left\{ \,
\begin{IEEEeqnarraybox}[][c]{l?s}
\IEEEstrut
PR_{i} & If task $i$ is running on machine $w$, \\
0 & Otherwise 
\IEEEstrut
\end{IEEEeqnarraybox}
\right.
\label{workePT}
\end{IEEEeqnarray} 

Here each task $i$ is being run on machine $w$ ($w = 1 ... m$) and we use equation \ref{index} to make unique indexes for tasks. In this way, we can access the processing throughput of each task in the implemented Storm system \ref{big_picture}.
\par 
Any mishandling of task assignment or taking an inappropriate number of instances for components may lead to over-utilization or under-utilization of CPUs on physical machines. Therefore, the adjustment of these two options to obtain maximum throughput without any machine overloading leads in large state space and extra computation complexity. Considering each task as an item which has specific profit $(PT_{iw})$ and weight $(TCU_{iw})$ and each machine as a knapsack with certain capacity $(MAC_w)$,  our problem can be mapped to the\textit{ multiple Knapsack optimization} problem [23]. The classic version of the Knapsack problem is NP-hard, so an approach to solve our problem is computationally infeasible.
\par 
There are some meta-heuristic algorithms to obtain the near-optimal answer for the \textit{multiple Knapsack problem} but most of them require a long time to give an answer. Whereas performing fast enough scheduling in stream data processing, is significantly important. For example, in case of machine failure, a slow scheduler leads the cluster to tuple overloading state. As well as any reconfiguration in the cluster needs new task assignment, so having a lazy scheduler is not acceptable. To derive a solution for the mentioned above problem, we have proposed a new heuristic algorithm that provides short execution time together with high throughput.

\section{Proposed Algorithm}

\subsection{Algorithm Overview}
\begin{figure}[!t]
	\centering
	\includegraphics[width=3.4in]{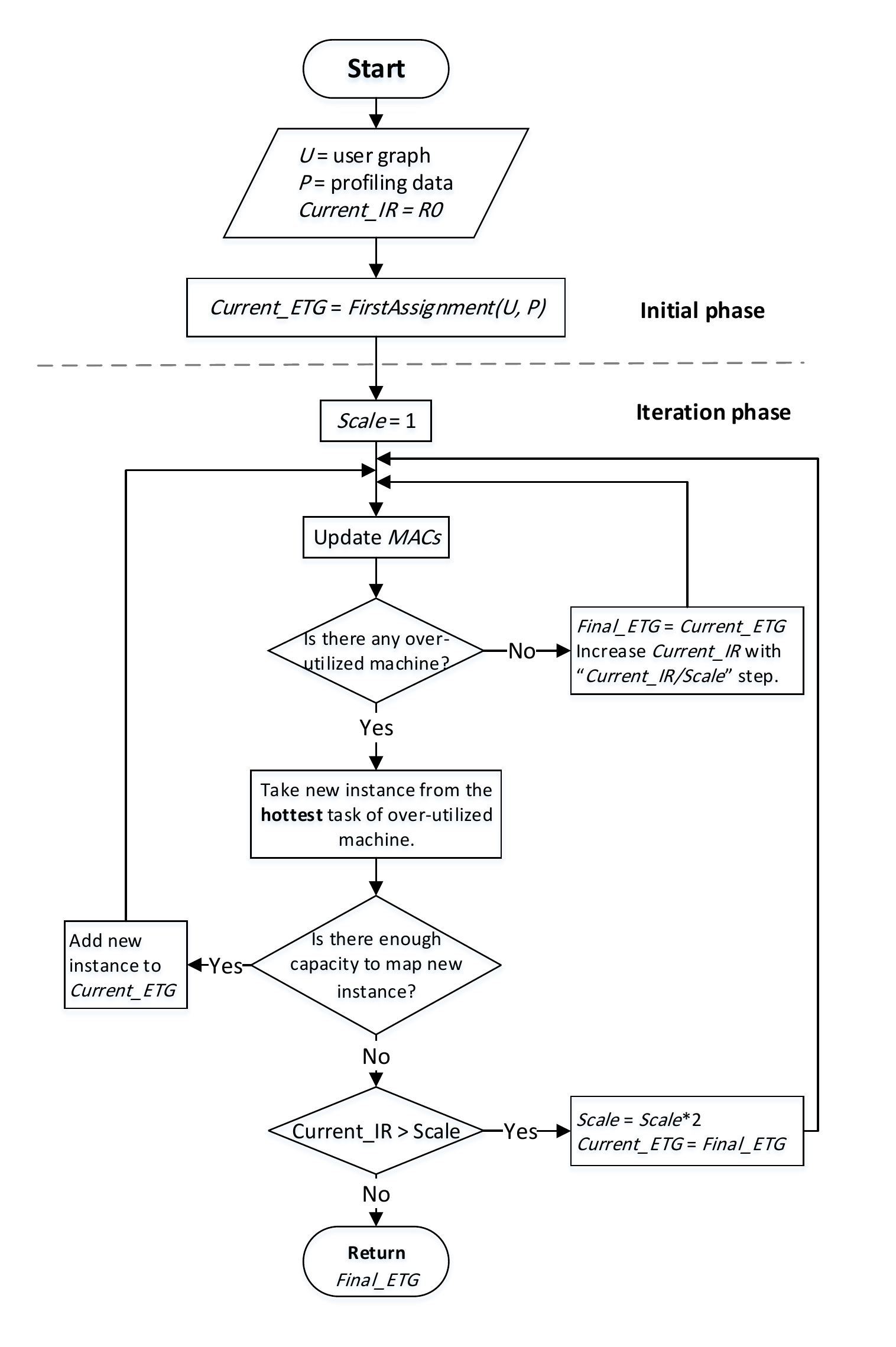}
	\caption{Proposed scheduling algorithm overview}
	\label{flowChart}
\end{figure}

Considering $UTG = (c_1, c_2, c_3, ...)$ as a user topology graph, we start to scale it up over a given cluster gradually, by increasing the topology input rate and taking more instances from the compute intensive components. We continue this process until almost fully utilize all nodes of the cluster. As Fig. \ref{flowChart} represents our scheduling algorithm has two phases: initial and iteration (separated by the dotted line). In the initial phase, we take a user topology graph and a set of $e_{ij}$s (profiling data), then take one instance from each component; finally, we map this primary execution graph to the cluster using \textit{FirstAssignment} procedure. In the iteration phase, we employ two options, including \textbf{increasing topology input rate} and \textbf{taking new instances}. The algorithm progressively uses these options in an iterative manner to achieve maximum throughput and fully utilize all of the available machines.
\par
At each iteration, \textit{MaximizeThroughput} procedure updates the \textit{MAC} values for all machines and check if any machine is over-utilized? When there is no over-utilized machine, it increases the topology input rate until at least one machine become over-utilized. In this situation, it takes a new instance from the component which one of its instances is bottleneck. Then it looks for the most suitable machine to place the instance on. If there is enough capacity to map the new instance, it is added to the execution topology graph $(ETG)$. But, if no candidate machine is found, the amount of input rate increment must be reduced. This procedure repeats this flow until the amount of input rate increment reaches its threshold and hence no more capacity is left on the cluster. Lastly, the iterations are finished and the final execution topology graph alongside with its corresponding mapping is obtained.
\par
In our algorithm, before mapping an instance to the proper machine, we need to know the amount of its CPU usage on all available types of machines. To have a good estimation of these values, we define some new metrics and present a heterogeneity-aware formula, which uses some profiling data to predict the $TCU$ of the intended instance on each machine. In the next sub-section, we explain this formula in detail.

\subsection{CPU Usage Prediction}
According to our experiments, for each task, the value of $TCU$ depends on its input rate, its computation complexity, and type of the machine, which the task is running on. Based on our observations, we assume the CPU usage of a task linearly increases in term of input rate increment. So we use (\ref{cpuPredict}) to estimate the $TCU$ of $i^{th}$ task on $j^{th}$ machine as follow:
\begin{IEEEeqnarray}{c}
	TCU_{ij} = (e_{ij} \times IR_i ) + MET_{ij}
	\label{cpuPredict}
\end{IEEEeqnarray}
\par
In this formula, $e_{ij}$ is the average execution time of $i^{th}$ task on $j^{th}$ machine to process a single tuple, $IR_i$ is the input rate of this task, and $MET_{ij}$ is miscellaneous execution time of Storm. According to our experiments, for each task $i$ running on machine $j$ the values of $e_{ij}$ and $MET_{ij}$ are independent of input rate, so they are considered as two constants.
\par
Our algorithm is designed on the basis of grasping the execution characteristics of different tasks. To know the performance characteristics of the individual types of nodes employed in the cluster for different types of tasks, we use a pre-process profiling [21]. In profiling, each task is run on all types of machines and each time we increase its input rate while it reaches the maximum value of its CPU usage. At this point, we measure the CPU usage, its corresponding input rate and the value of $e_{ij}$ which can be measured using \textit{get\_execute\_ms\_avg()} Java function or Storm's UI. Knowing $TCU$, $IR$ and $e_{ij}$, the value of $MET_{ij}$ is calculated using (\ref{cpuPredict}). The reasoning behind choosing the maximum point is that the variation in measured CPU utilization is very low when the processing load is either relatively low or relatively high [14]. As the number of arrival tuples is increased, the portion of the miscellaneous overhead of Storm and CPU idle time are decreased; thus, the prediction of CPU usage at the maximum point is quite accurate. 
\par
After profiling, we have a constant $e_{ij}$ and $MET_{ij}$ for all possible assignments of topology's components to cluster's nodes, thus we are able to predict its $TCU$ for each arbitrary value of $IR$. 
\par
As topology input rate has a domino effect on all tasks, we define a new metric (called $\alpha$), to estimate the output rate of each task. For task $i$, $\alpha_i$ is a given parameter (extracted from profiling data), that shows the average ratio of its output tuples count to its input tuples count. Now we can calculate the input rate of each downstream bolt, using equation \ref{inputRate} and then obtain its corresponding CPU usage by formula \ref{cpuPredict}. 
\begin{IEEEeqnarray}{c}
	IR_{next\_stage\_task} = \sum_{i=1}^{y} (\frac{OR_i}{x}) 
	\label{inputRate}
\end{IEEEeqnarray}
where
\begin{IEEEeqnarray}{c}
	OR_i = IR_i \times \alpha_i   \nonumber
	\label{outputRate}
\end{IEEEeqnarray}
\par
In this equation, $x$ is the number of downstream tasks which are fed by $i^{th}$ task, and $y$ is the number of tasks which are feeding the intended downstream task \textit{(next stage task)}. Here, $OR_i$  is output rate of  $i^{th}$ task, and $IR_i$ is its input rate. 

\subsection{FirstAssignment}
To meet the first constraint of our problem, we use \textit{FirstAssignment} procedure to take one instance from each component of the given topology. This procedure takes a user topology graph \textit{UTG}  and a set of $e_{ij}$s and $MET_{ij}$ (profiling data) as input, according to Algorithm \ref{firstAssign}. For a given topology with an initial input rate $R_0$, it estimates the input rate of each component, using $\alpha$ values and equation \ref{inputRate}. Then, for each component, it predicts the \textit{TCU} on different machines using profiling data and formula \ref{cpuPredict} and assign the component to the machine which results least \textit{TCU}. In the case of a successful assignment, one instance of the component is added to the temporary execution topology (\textit{Current\_ETG}). After mapping all the components, a primary execution topology with a specified task assignment is given to the \textit{MaximizeThroughput} procedure. 
\begin{algorithm}
	\SetKwInOut{Input}{Input}
	\SetKwInOut{Output}{Output}
	\SetKwInOut{Define}{Define}	
	\Input{User graph $U$, Profiling information $P$}
	\Output{Assigned user graph $Current\_ETG$ to the target cluster}
	$Current\_IR = R_0$\\
	$Current\_ETG = Null $ \\
	\ForEach{component $i$ in $U$}{
		Using P map $i$ on the machine which results the least $TCU$ \\
		Add $i$ to $Current\_ETG$
	}
	\caption{First Assignment}
	\label{firstAssign}
\end{algorithm}

\subsection{MaximizeThroughput}
In the second phase we use a progressive algorithm to use all processing power of the cluster nodes efficiently, and hence maximize the overall throughput. To do that another procedure called \textit{MaximizeThroughput}, takes profiling data and a primary execution topology graph from the \textit{FirstAssignment} procedure, as inputs. After a limited number of iterations, this procedure results in a final execution topology graph (\textit{Final\_ETG}) in which it determines how many instances each component has, where each instance must be placed on the target cluster and the maximum value of topology input rate which cluster can tolerate. 
\par
Algorithm \ref{secondAssign} represents pseudo-code of \textit{MaximizeThroughput} procedure. Here, variable \textit{Scale} is defined to control the amount of input rate increment, which its value is initialized to be 1. At the beginning of the algorithm, we assume \textit{Current\_ETG} is mapped to the cluster and its input rate starts from $R_0$. In the next step, the algorithm starts to increase the input rate, progressively and at each iteration, it updates the value of \textit{MAC} for all machines using formula \ref{cpuPredict}. After that, it checks if any machine is over-utilized. Now there are two possibilities: 
\\
\textbf{1- No machine is over-utilized}; in this situation current execution topology graph (\textit{Current\_ETG}) and its corresponding input rate are retained in \textit{Final\_ETG}  as the latest stable state; and then topology input rate is increased by a factor proportional to \textit{Currect\_IR/Scale}. By increment of topology input rate, the processing load of all the tasks will be increased, thus their utilization on the corresponding machine needs to be updated, so we return to line 1 to evaluate new conditions. 
\textbf{2- At least one machine is over-utilized}; in this situation, we take a new instance from component corresponding to the task with the highest CPU usage (\textit{hottest} task) in the first over-utilized machine. This new instance should be assigned to one of the available machines so we look for the most suitable machine to map this instance on. When there is at least one machine with enough capacity to serve this new instance, it is added to the \textit{Current\_ETG} alongside with its assignment information; then we go to line 1 to update the $MAC$s. But, if no candidate machine is found to serve that, the termination condition is checked. By taking a new instance from $i^{th}$ component a portion of the upstream data ($IR_i$) will be processed by this instance and hence the input rate of other instances including the \textit{hottest} task is decreased, so over-utilization is solved.
\par
While current input rate (\textit{Current\_IR}) is greater than value of \textit{Scale}: I) The amount of input rate increment will be decreased by a factor proportional to $\frac{1}{2}$, in other words the variable \textit{Scale} is duplicated. II) The latest stable state of scheduling is recovered by copying the \textit{Final\_ETG} to the \textit{Current\_ETG}, because the current rate exceeds the cluster processing capacity. III) Return to line 1 to update the value of $MAC$s. 
\begin{algorithm}
	\SetKwInOut{Input}{Input}
	\SetKwInOut{Output}{Output}
	\Input{User graph $Current\_ETG$, Profiling information $P$}
	\Output{$Final\_ET$ with its mapping information on target cluster}
	Update $MACs$ using CPU prediction formula and $P$
	\eIf{no CPU over-utilization}{
		$Final\_ETG = Current\_ETG$	\\
		$Current\_IR \hspace{0.2cm} += \hspace{0.2cm} Current\_IR / Scale$ \\
		\textbf{go to} line 1
	}
	{
		Take new instance $h$ from $hottest$ task \\
		\eIf{enough capacity exists on cluster}
		{
			Add $h$ to $Current\_ETG$\\
			\textbf{go to} line 1
		}
		{
			\eIf{Current\_IR $>$ Scale}
			{
				$Scale = 2*Scale$ \\
				$Current\_ETG = Final\_ETG$ \\
				\textbf{go to} line 1
			}
			{
				\textbf{return}  $Final\_ETG$
			}	
		}
	}
	\caption{Maximize Throughput}
	\label{secondAssign}
\end{algorithm}

\par
In this procedure, we regulate the incrementation of input rate in lines 3 and 12; and use \textbf{take new instance} option in lines 6-8. This process is repeated until the termination condition (\textit{Current\_IR $ \leq$ Scale}) is satisfied; it means the input rate cannot be increased anymore and no capacity is left in the cluster to place any new instance. Now all machines are almost fully utilized, and the final execution graph alongside with its corresponding assignment is created, so the algorithm ends.

\section{Evaluation} 

\subsection{Experimental Setup}
To evaluate the proposed scheduler, we have implemented it in Storm as a new scheduler [24]. Using this scheduler we run benchmark topologies on a cluster of four heterogeneous machines. Table \ref{tblSysChar} shows the specifications of cluster nodes. In our cluster, all machines are connected together through a switch with 1 Gb/s network adapters. One of the Core i3 machines is used as a master node, which runs Zookeeper and Nimbus daemons, and other machines are configured as worker nodes. All experiments are obtained with Apache Storm 0.9.5 installed on top of Ubuntu 16.04.
\begin{table}[!t]
	\renewcommand{\arraystretch}{1.2}
	\caption{System characteristics}
	\label{tblSysChar}
	\centering
	\begin{tabular}{p{0.52in} p{0.40in} p{0.75in} p{0.42in} p{0.45in}}
		\hline
		\bfseries Cluster Node & \bfseries Memory & \bfseries Processor & \bfseries Network adapter & \bfseries Operating System\\
		\hline
		Machine 1 & 2GB & Pentium Dual-Core 2.6 GHz & 1 Gb/s & Ubuntu 16.04 \\
		Machine 2 & 4 GB & Intel Core i3 2.9 GHz & 1 Gb/s & Ubuntu 16.04 \\
		Machine 3 & 6 GB & Intel Core i5 2.5 GHz & 1 Gb/s & Ubuntu 16.04 \\
		Machine 4 & 4 GB & Intel Core i3 2.9 GHz & 1 Gb/s & Ubuntu 16.04 \\
		\hline
	\end{tabular}
\end{table}
\par
\begin{figure}[!t]
	\centering
	\includegraphics[width=2.7in]{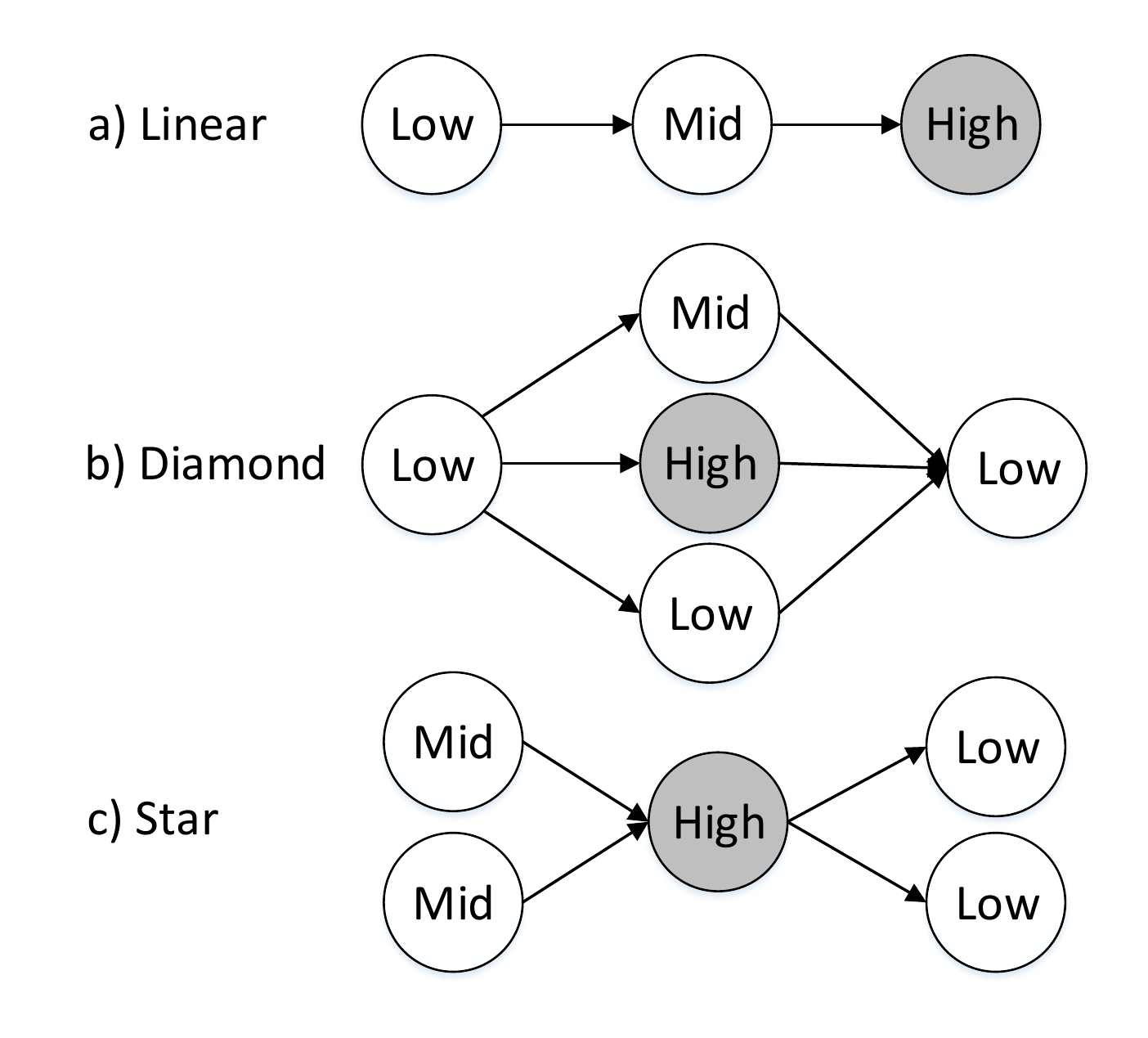}
	\caption{Layout of Micro-benchmark[6] topologies}
	\label{topologies}
\end{figure}

However, Storm topologies can be an arbitrary DAG, most of the topologies are a mixture of three basic topologies: Linear, Diamond, and Star (Fig. \ref{topologies}).  \textit{Linear} topology has one source, a sequence of intermediate components and a sink. When there are several parallel components between the source and sink it is known as \textit{Diamond} topology and a \textit{Star} topology has multiple sources connected to a single component and this intermediate component is the parent of multiple sinks. There are some production applications from industry like \textit{PageLoad} topology, \textit{Processing topology}, and \textit{Network Monitoring topology} [7], which has been used in [6], [7], [9] to evaluate their works. All these topologies are a combination of Linear, Diamond, and Star topologies, and are good choices where the processing power, network capacity, and topology connectivity must be considered. Whereas in this work we only focus on the required processing power by each component and have no network consideration; therefore, we used three basic topologies from Micro-Benchmark [6] to evaluate the effectiveness of our scheduling algorithm. All topologies in this benchmark are made from three types of CPU intensive components called \textit{lowCompute}, \textit{midCompute}, and \textit{highCompute}. Table \ref{tblProfiling} shows the profiling data $e_{ij}$ of each component, on all types of machines which mentioned in Table \ref{tblSysChar}.
\begin{table}[!t]
	\renewcommand{\arraystretch}{1.3}
	\caption{Profiling of topologies' tasks on cluster's machines}
	\label{tblProfiling}
	\centering
	\begin{tabular}{c c c c}
		\bfseries Task Type & \bfseries Machine 1 & \bfseries Machine 2 & \bfseries Machine 3 \\
		\hline
		lowCompute & 0.0581 (s) & 0.107 (s)& 0.0916 (s)\\
		midCompute & 0.103 (s)& 0.1844 (s)& 0.168 (s)\\
		highCompute & 0.1915 (s)& 0.3449 (s)& 0.3207 (s)\\
	\end{tabular}
\end{table}

\par
To maximize the throughput of the system, the algorithm tries to make a fitting execution topology graph for a specific heterogeneous cluster. Since finding the optimum number of instances for each component has a very large design space, we used two simple topologies from Storm-Benchmark [15]. They both contain two components and make it possible to verify how well the algorithm calculates the number of instances for each component. As there is no similar scheduling algorithm to produce the execution graph according to the computing power of heterogeneous machines, we compared its results with optimal execution graph. 

\subsection{Experimental Evaluation}
\begin{large}\fontsize{11}{1in}\textbf{Evaluation of CPU usage formula:}\end{large} The first part of our proposed algorithm, which needs to be verified, is the CPU usage prediction formula. An interesting outcome of the empirical case studies is that the difference between measured $TCU$ and its calculated value is very low whenever the input rate is either relatively low or relatively high. Thus, the prediction of CPU usage is quite accurate when the CPU is either lightly or heavily loaded. Our experiments show when the CPU usage is moderate, the measured $TCU$ has more variation than its calculated value and thus is less predictable. However, even the largest difference between measured and predicted $TCU$ always was less than 8\%.
\par
Our experiments in this part, cover different conditions for a unique bolt, including all possible combinations of three types of processors and three different structures of topologies. For each experiment the \textit{highCompute} bolt (gray bolt in Fig. \ref{topologies}) is placed on a single machine and its upstream bolts are placed on the powerful enough machines, such that they were able to fully utilize it. For example, in the first experiment \textit{highCompute} bolt of Linear topology is assigned to the machine with Pentium processor and its real CPU usage is measured when the input rate of topology is 8 tuple per second. Then the input rate is increased by a factor of a random number between 20 and 80, and new CPU usage is measured again. This process is continued while the processor becomes over-utilized, so the input rate cannot be increased any more. 
\par
\begin{figure*}[!t]
	\centering
	\subfloat[Linear topology]{\includegraphics[width=2.3in]{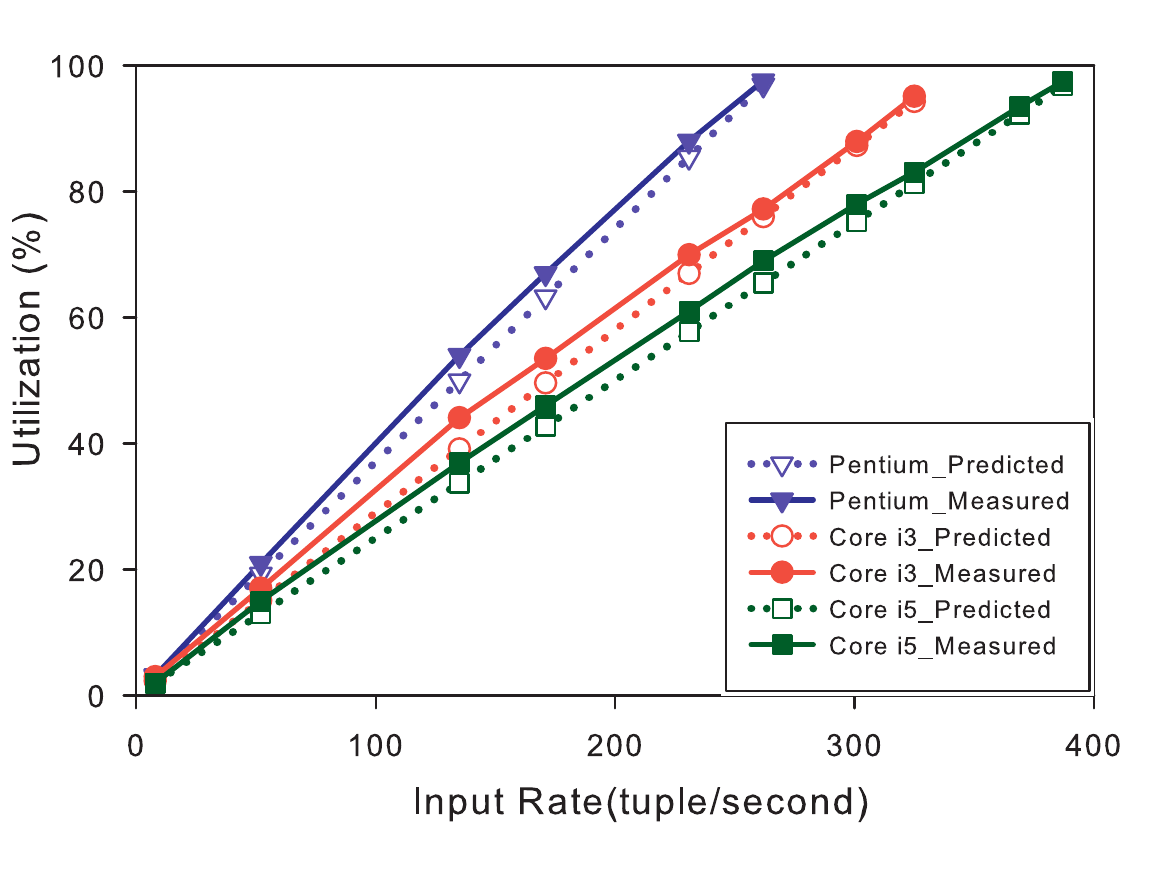}
	\label{cpu_pred_linear}}
	\hfil
	\subfloat[Diamond topology]{\includegraphics[width=2.3in]{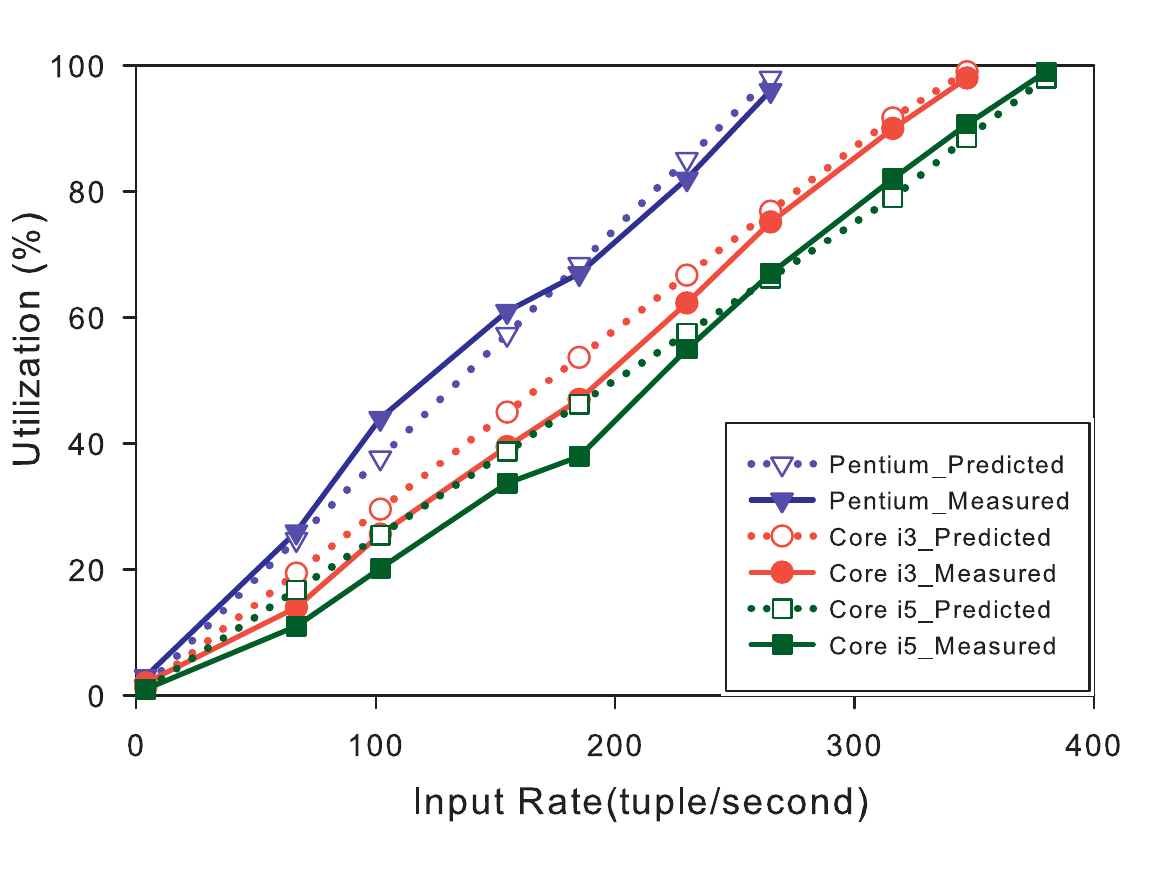}
	\label{cpu_pred_diamond}}
	\hfil
	\subfloat[Star topology]{\includegraphics[width=2.3in]{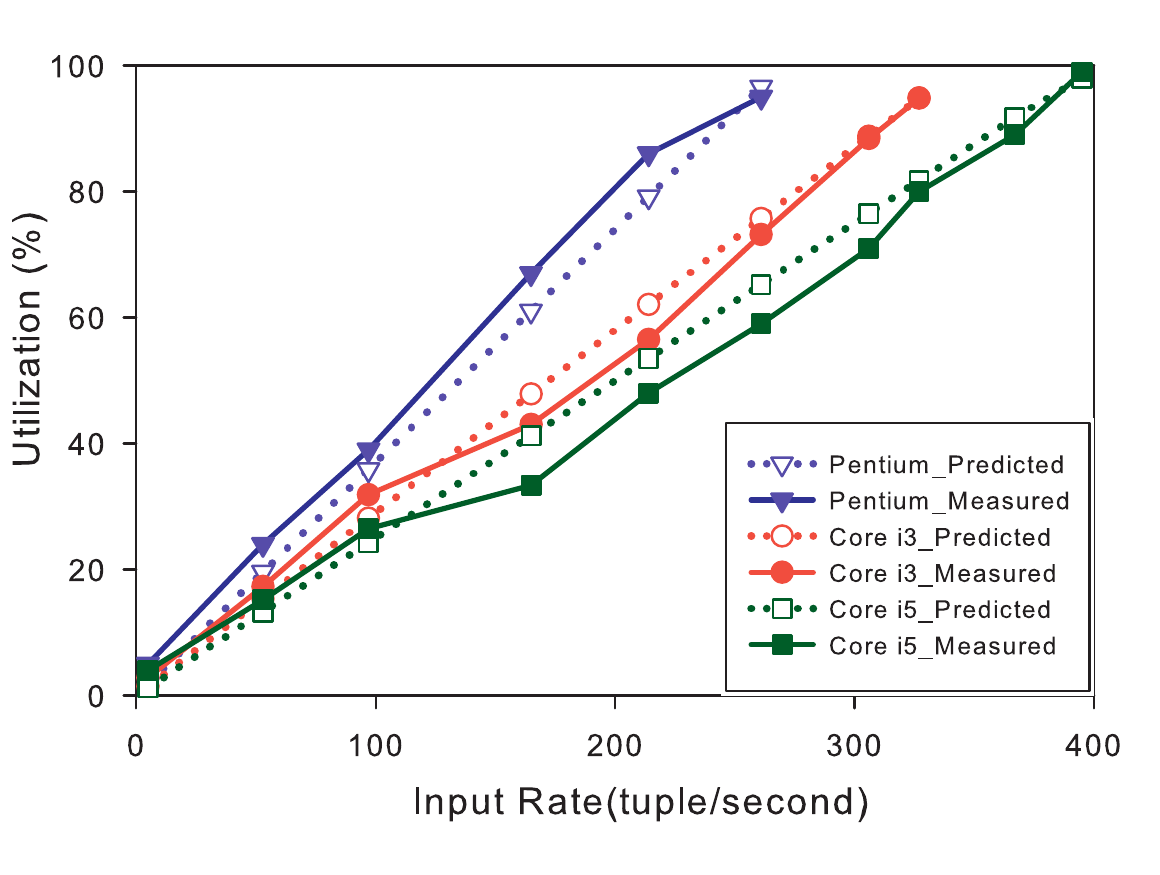}
	\label{cpu_pred_star}}
	\caption{Predicted and measured CPU utilization comparison for highCompute bolt on different machines}
	\label{cpu_prediction}
\end{figure*}

Fig. \ref{cpu_prediction} represents the measured and predicted $TCU$ of \textit{highCompute} bolt for different values of input rate, on three different types of CPUs for Linear, Diamond and Star topologies respectively. Here the unfilled dotted lines show the predicted value of $TCU$ that is obtained by formula \ref{cpuPredict} and the filled dotted lines show real $TCU$, measured by collector tool [16].

\begin{figure*}[!t]
	\centering
	\includegraphics[width=7in]{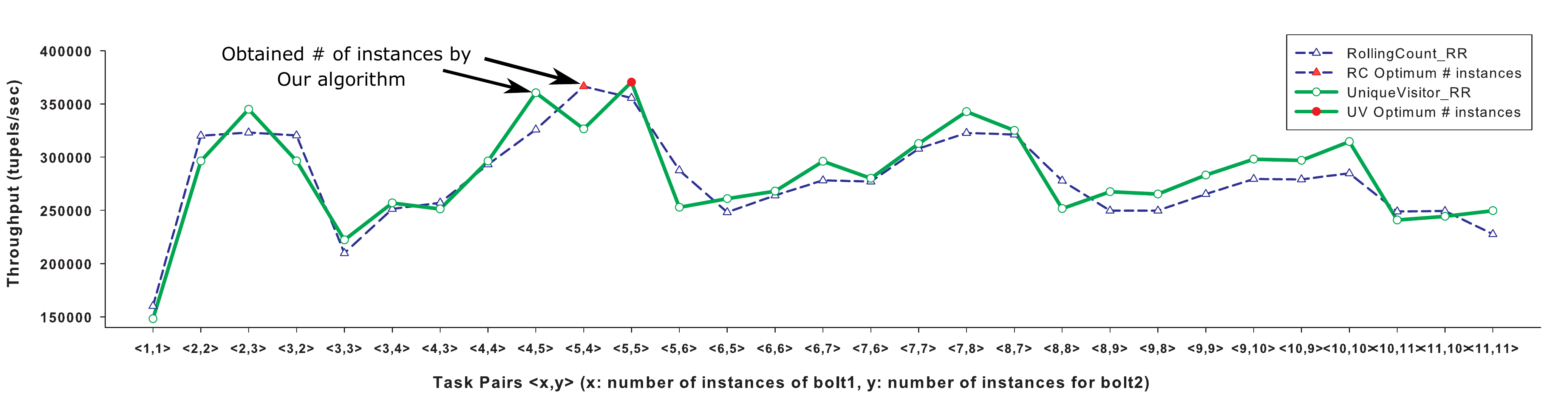}
	\caption{Maximum achievable overall throughput in terms of number of instances}
	\label{exec_chart}
\end{figure*}

\begin{large}\setlength{\parindent}{0em}
\setlength{\parskip}{0.5em}\fontsize{11}{1in}\textbf{Evaluation of execution graph:}\end{large} One outcome of our scheduling algorithm is obtaining the near-optimal number of instances for each component. Fig. \ref{exec_chart} depicts overall throughput by different number of instances for both \textit{RollingCount} and \textit{UniqueVisitor} topologies from Storm-Benchmark [15]. The $<$\textit{x,y}$>$ pairs on the horizontal axis show the number of instances for bolt1 and bolt2 respectively in both topologies. To observe the effect of the structure of the execution graph on overall throughput, different execution graphs are scheduled using default scheduler of Storm. Then our algorithm is performed to find out how well it calculates the number of instances for each topology. Our algorithm obtained pair $<$\textit{5,4}$>$ for \textit{RollingCount} topology, which exactly is the optimal number of instances. It also obtained pair $<$\textit{4,5}$>$ for \textit{UniqueVisitor} topology, which has the closest number of instances to the optimal pair $<$\textit{5,5}$>$, and it eventuates only 2\% throughput decrement than optimal throughput. The obtained pairs by our algorithm for both topologies are shown by the arrow in Fig. \ref{exec_chart}.

\begin{figure*}[!t]
	\centering
	\subfloat[Linear topology]{\includegraphics[width=2.3in]{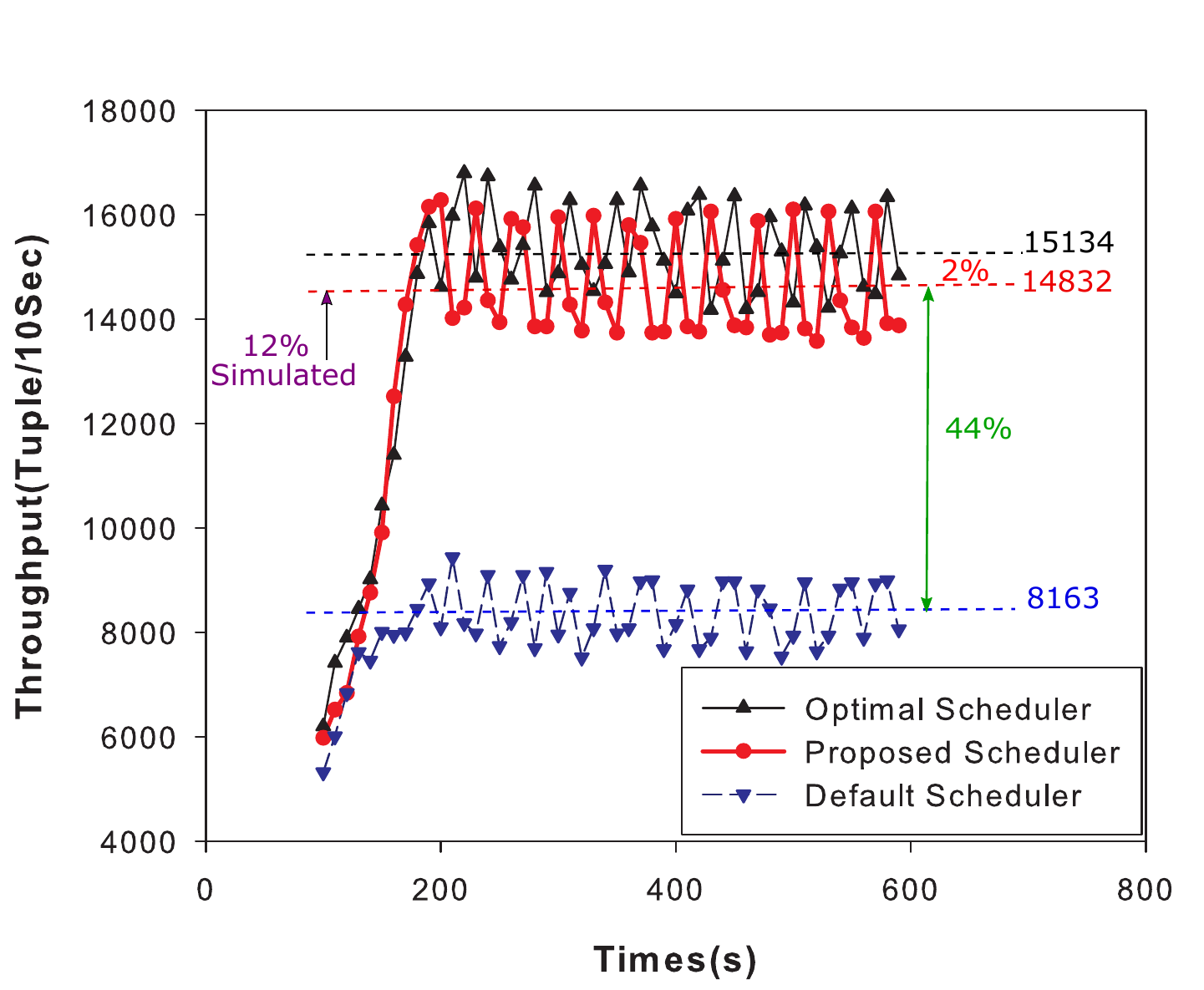}
		\label{exp_linear}}
	\hfil
	\subfloat[Diamond topology]{\includegraphics[width=2.3in]{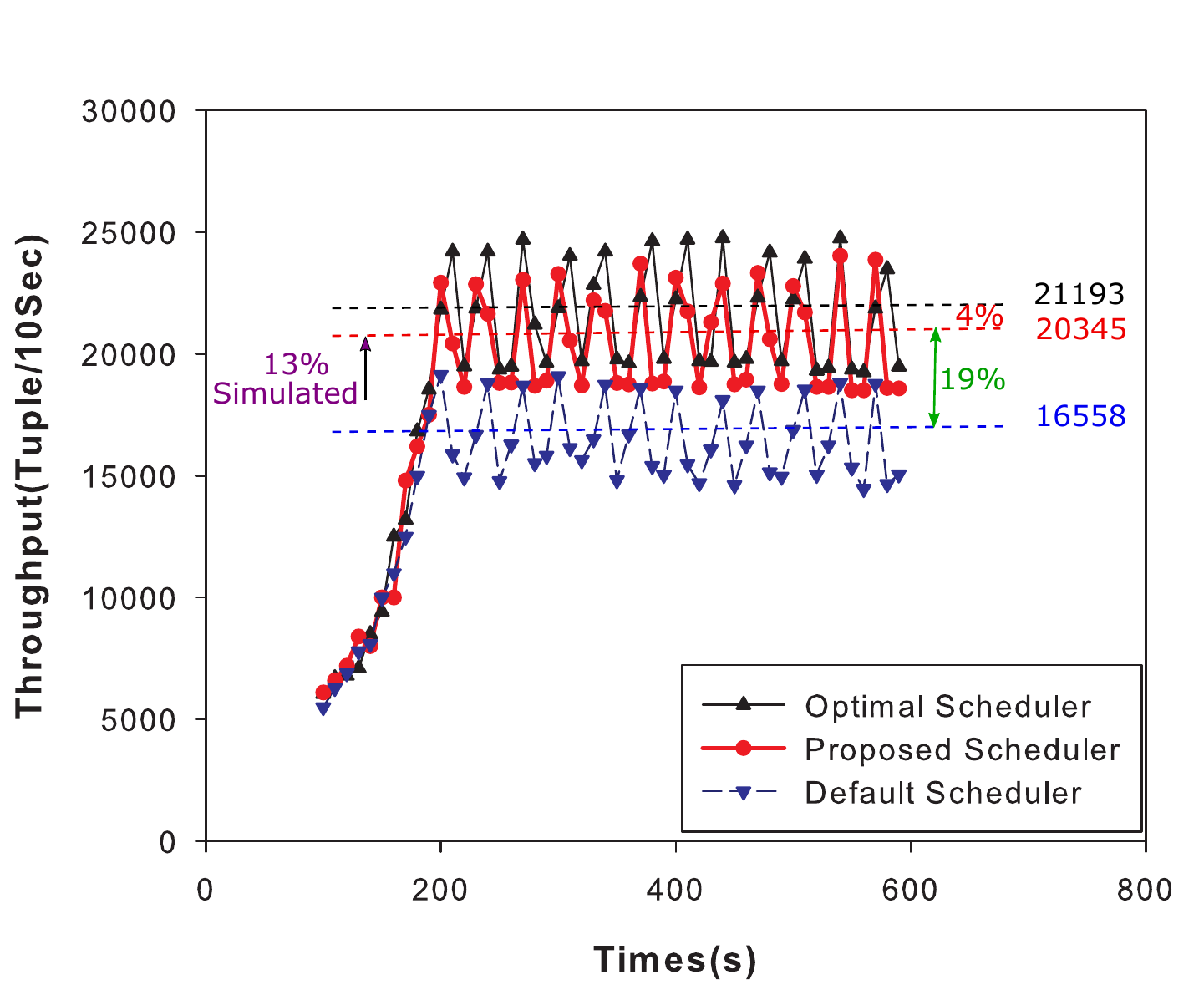}
		\label{exp_diamond}}
	\hfil
	\subfloat[Star topology]{\includegraphics[width=2.3in]{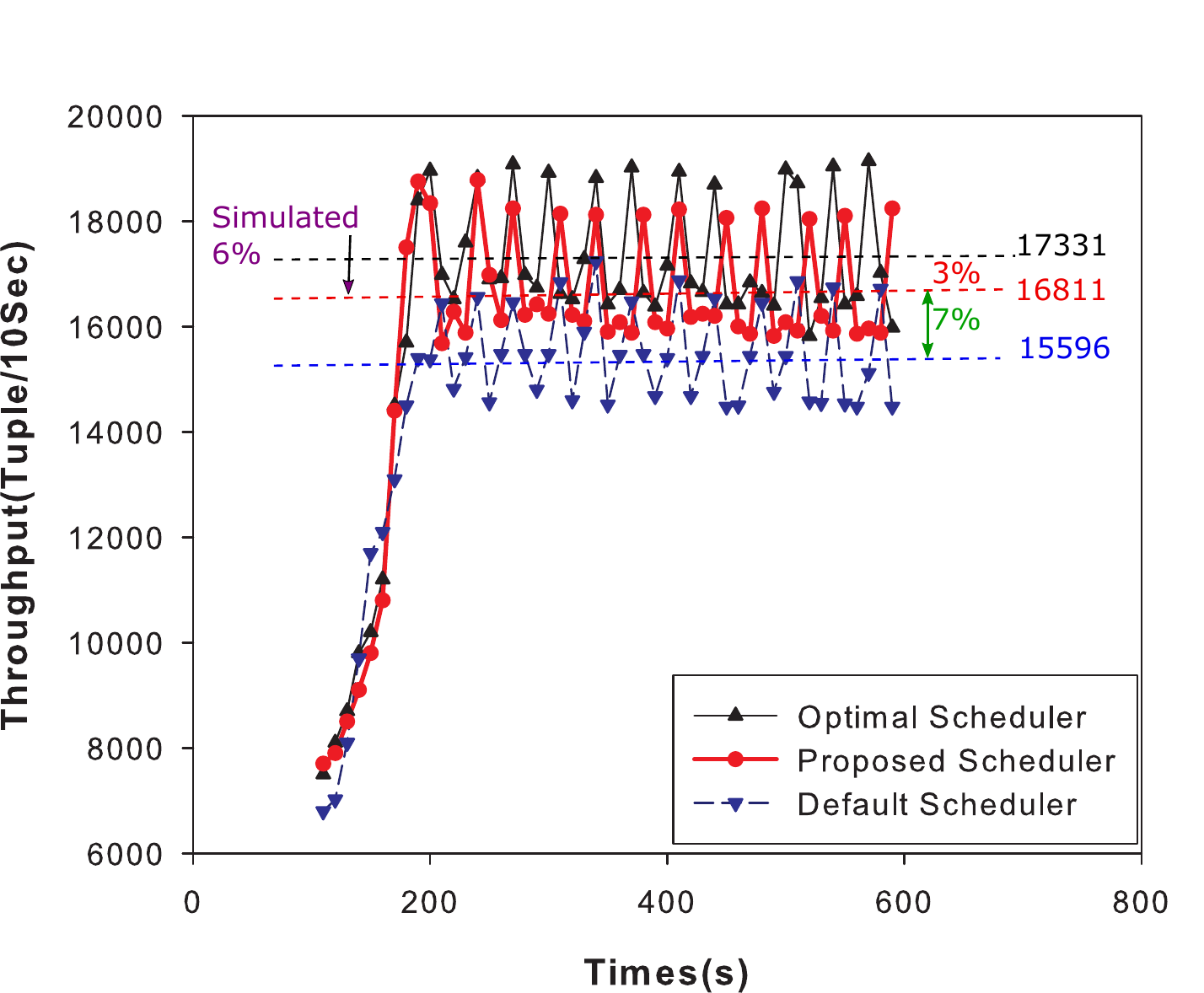}
		\label{exp_star}}
	\caption{Experimental throughput comparison of default, proposed and optimal schedulers for different topologies}
	\label{exp_results}
\end{figure*}

\begin{large}\setlength{\parindent}{0em}
\setlength{\parskip}{0.5em}\fontsize{11}{1in}\textbf{Throughput Comparison:}\end{large} We evaluate the performance of the proposed algorithm in term of throughput. Here the efficiency of its task assignment is compared with Storm's default scheduler and the optimal scheduler. Fig. \ref{exp_results} depicts the experimental results of our scheduler in comparison with other schedulers. According to this figure, our scheduling method provides 7\% to 44\% throughput enhancement in comparison with Storm's default scheduler while it can find the solution within 4\% of the throughput of the optimal scheduler in the worst case (Fig. \ref{exp_diamond}). 
\par
As we can see, for Linear topology our scheduler results in 44\% higher throughput than default scheduler, while it is 2\% less than optimal throughput, but for other topologies, it provides less gain. This is due to the different characteristics of the target topologies. In some topologies such as Linear, the main challenge is the processing power, so the proposed scheduler would result in almost best achievable throughput but when other resources such as network or memory get bottleneck, we need to consider several factors to obtain a better result. 

\begin{figure*}[!t]
	\centering
	\includegraphics[width=6in]{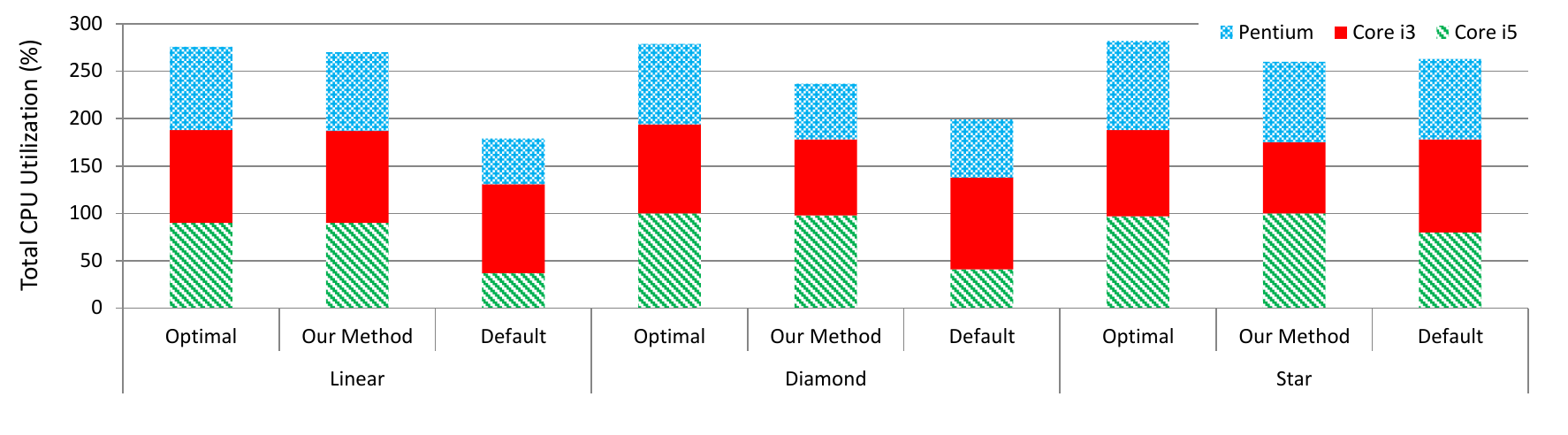}
	\caption{Obtained CPU utilizations of worker nodes by different scheduler for Micro-benchmark topologies}
	\label{util_chart}
\end{figure*}

\begin{large}\setlength{\parindent}{0em}
\setlength{\parskip}{0.5em}\fontsize{11}{1in}\textbf{Utilization Comparison:}\end{large} Normally better resource utilization results in higher throughput, but to make sure that our scheduler has efficient CPU usage, we need to compare CPU utilization of all schedulers. In Fig. \ref{util_chart} we can see the total CPU utilization of cluster nodes for Linear, Diamond, and Star topologies, under different scheduling methods. In all cases, the optimal scheduler uses the CPUs efficiently, so it has the highest summation of CPU utilization percentages. 
\par
When Diamond and Linear topologies are running on the cluster, our scheduler has more CPU usage, therefore higher throughput than default scheduler. But in the case of Star topology, its total CPU usage is less than default scheduler, although it still has higher throughput. The reason is that our scheduler uses the most powerful processor (Core i5) better than default scheduler. Therefore, it always uses the processing resources more efficiently than the default scheduler.

\subsection{Large-Scale Clusters Evaluation}
\begin{large}\setlength{\parindent}{0em}
\setlength{\parskip}{0.5em}\fontsize{11}{1in}\textbf{System Simulation:}\end{large} Due to resource limitation and real requirements of big data applications we need to use a modeling scenario. Generally, simulation makes it possible to model bigger distributed stream processing systems. To be able to simulate the real use case scenario in which Apache Storm is being used, we needed to simulate real heterogeneous environments. For each specific scheduling policy, our simulation scenario takes the processing characteristics of both cluster nodes and topology components, and then it reports the overall throughput of the topology and CPU utilization of existing machines [25]. 
\par
To make sure that our simulator has enough accuracy, we performed it in the same conditions of our real experiments. As shown in Fig. \ref{exp_results} the difference between implementation and simulation results is less than 13\% in the worst case, which means the simulator has acceptable accuracy, so we can use this simulator to evaluate our scheduler in case of large scale clusters.

\begin{large}\setlength{\parindent}{0em}
\setlength{\parskip}{0.5em}\fontsize{11}{1in}\textbf{Simulation Results:}\end{large} For each topology (Linear, Diamond, and Star), we need to find an appropriate execution graph according to the specification of cluster nodes. Therefore, we first run our algorithm to determine the number of instances for each component for the intended cluster. Now we can fairly compare only the effectiveness of scheduling policies (proposed and default schedulers) in terms of overall throughput and resource utilization. These simulations are executed for three different clusters, with the combinations of heterogeneous machines, shown in Table \ref{tblScenarios}.
\begin{table*}[!t]
	\renewcommand{\arraystretch}{1.3}
	\caption{Number of machines and component's instances of considered scenarios}
	\label{tblScenarios}
	\centering
	\begin{tabular}{|P{0.37in}|P{0.38in}|P{0.8in}|P{0.8in}|P{0.8in}|c|c|c|c|c|c|c|c|c|}
		\hline
		 & &  &  &  & \multicolumn{9}{c|}{\bfseries Tasks} \\
		\bfseries Scenario & \bfseries Cluster Type &\bfseries $\#$ of Machine 1 & \bfseries $\#$ of Machine 2  & \bfseries $\#$ of Machine 3& \multicolumn{3}{c|}{\bfseries Linear} & \multicolumn{3}{c|}{\bfseries Diamond} & \multicolumn{3}{c|}{\bfseries Star}\\
		& & & & & L & M & H & L & M & H & L & M & H \\
		\hline
		1 & Small & 2 & 2 & 2 & 3 & 3 & 7 & 12 & 5 & 7 & 9 & 10 & 5 \\
		2 & Medium & 10 & 10 & 10 & 41 & 42 & 35 & 67 & 15 & 46 & 45 & 34 & 25\\
		3 & Large & 20 & 70 & 90 & 201 & 156 & 341 & 397& 208 & 91 & 327 & 156 & 206\\
		\hline
	\end{tabular}
\end{table*}

\par
Our obtained simulation results report both throughput and utilization, per each node of the intended cluster. We calculate the overall throughput of topology by adding together these throughput values; and for utilization, we calculate a weighted average of reported utilization, with weights determined by equation \ref{equUtil}. Because of cluster heterogeneity, it is important to give more value to the machines with more processing capacity. Therefore, first, we determine the weight of each machine using profiling data according to equation \ref{equWeight}, then calculate overall utilization, using equation \ref{equUtil}.
\begin{IEEEeqnarray}{c}
	U = \sum_{i=1}^{T} x_i \bar{u_i} 
	\label{equUtil}
\end{IEEEeqnarray}
\par
In this equation, $U$ is the overall utilization of topology on the target cluster with $T$ types of machines. Here $\bar{u_i}$  is the average CPU utilization of all machines of type \textit{i} and $x_i$ is calculated as follow:
\begin{IEEEeqnarray}{c}
	x_i = \sum_{j=1}^{C} x_{ij}   
	\label{equWeight}
\end{IEEEeqnarray}
where
\begin{IEEEeqnarray}{c}
	x_{ij} = \frac{\frac{1}{e_{ij}}}{\sum_{k=1}^{T} (\frac{1}{e_{ik}})} \nonumber
	\label{equWeightI}
\end{IEEEeqnarray}
\par		
For each type of machines, we use equation \ref{equWeight} to determine its weight. Variable $C$ shows how many types of components does the topology have; it is different from $n$ which shows the total number of components of topology, because a topology may have several components with the same type ($C \leq n $).
\par
\begin{figure*}[!t]
	\centering
	\subfloat[Scenario 1 (small cluster)]{\includegraphics[width=2.29in]{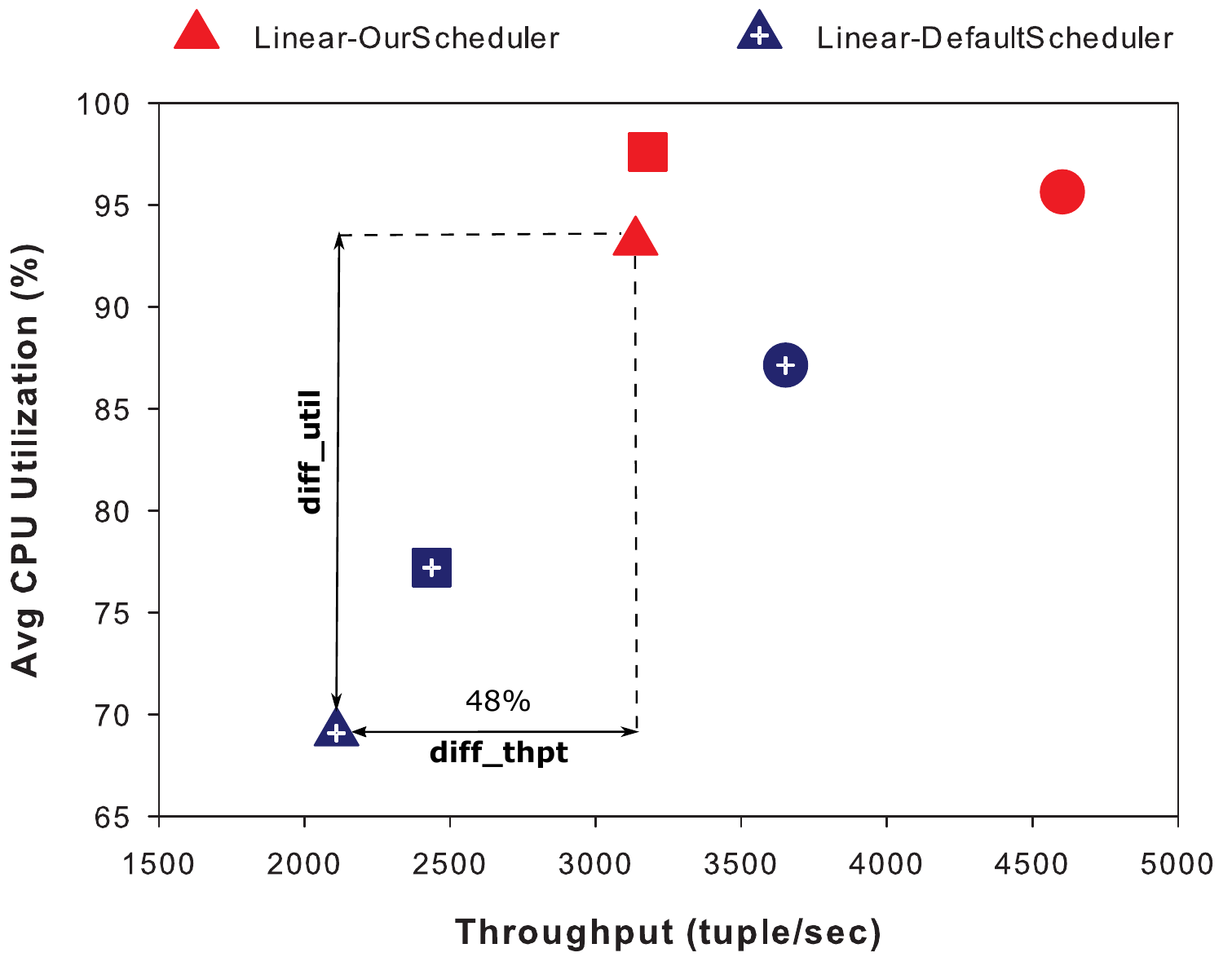}
		\label{sim_scen1}}
	\hfil
	\subfloat[Scenario 2 (medium cluster)]{\includegraphics[width=2.32in]{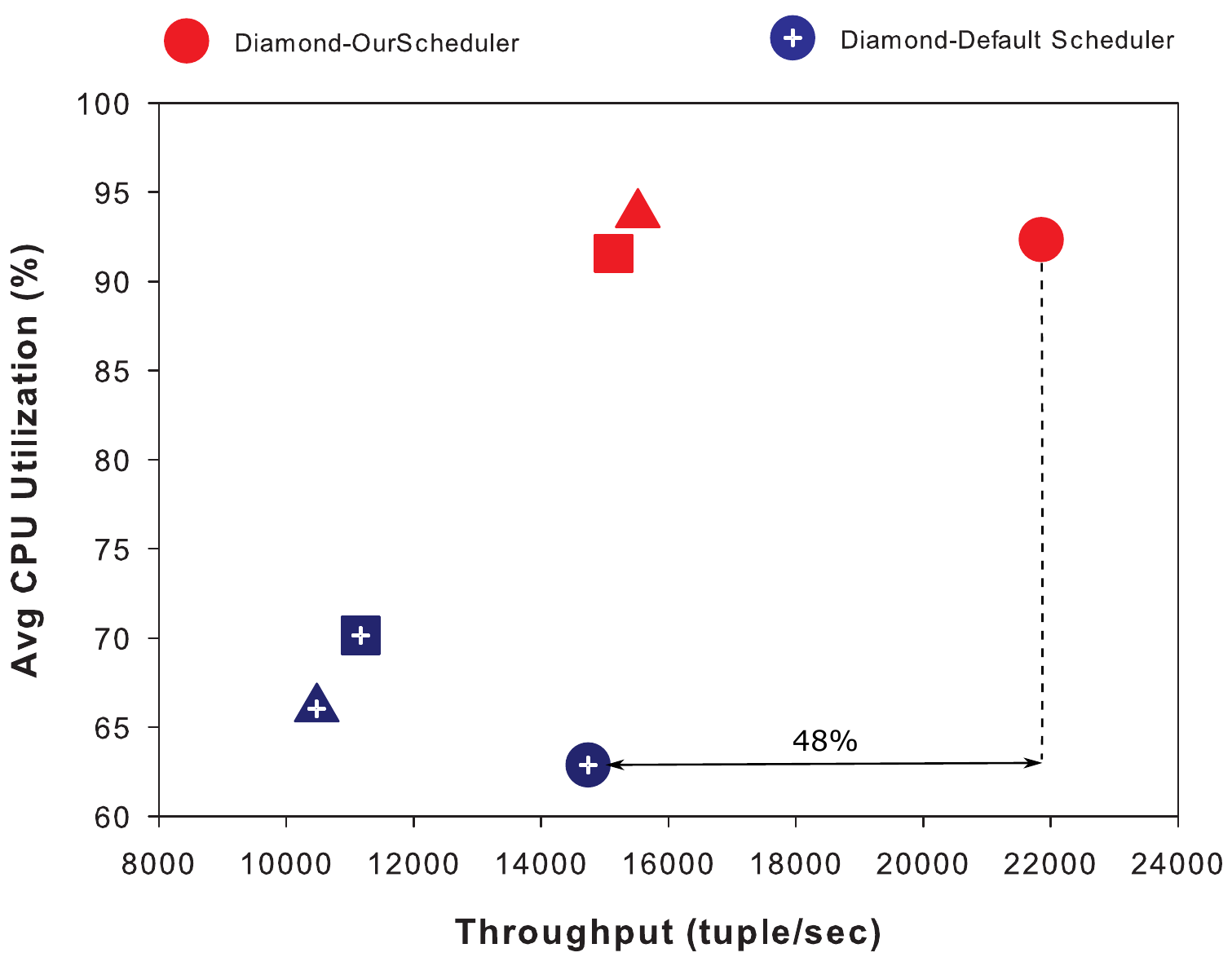}
		\label{sim_scen2}}
	\hfil
	\subfloat[Scenario 3 (large cluster)]{\includegraphics[width=2.29in]{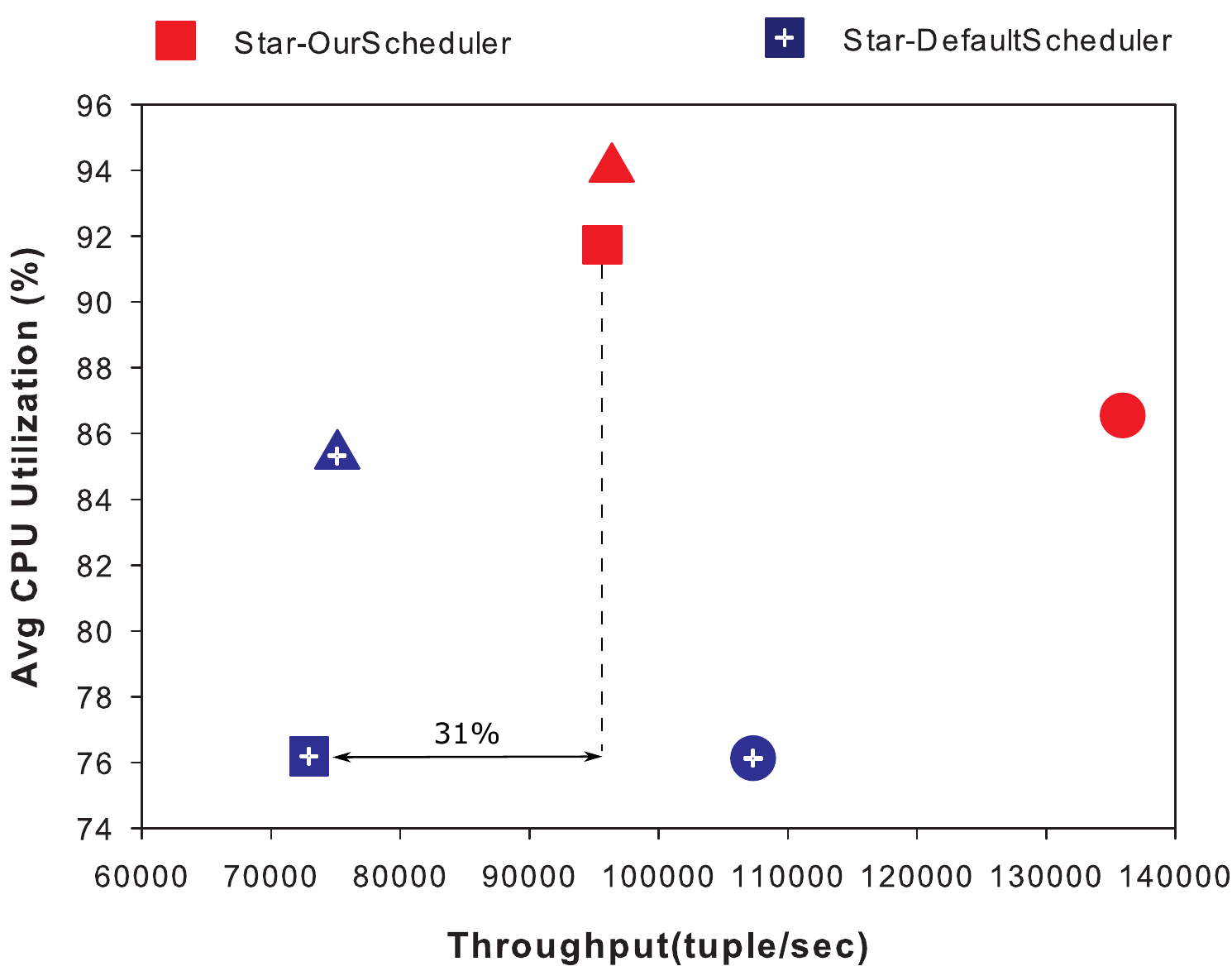}
		\label{sim_scen3}}
	\caption{Simulation results comparison of default and proposed schedulers for different topologies in considered scenarios}
	\label{sim_scenarios}
\end{figure*}

Fig. \ref{sim_scenarios} shows both throughput and utilization comparison of the proposed and default scheduler for three different topologies. According to these simulating results, for the small cluster (scenario 1) our scheduler has 26\% to 49\% throughput gain compared to default scheduler while it better utilizes the processing resources 10\% to 35\%. In the medium cluster (scenario 2) the proposed scheduler eventuates 36\% to 48\% improvement for overall throughput and 31\% to 47\% more resource efficiency. Finally, for the large cluster (scenario 3) the proposed scheduler in comparison with default scheduler provides 27\% to 31\% and 10\% to 21\% gain for throughput and utilization respectively. 
\par
Note that higher CPU utilization by itself is not an advantage for our scheduler, but when the ratio of differences of throughputs to utilizations between our scheduler and default scheduler is greater than 1, it can be considered as an advantage. It shows how efficient our proposed scheduler uses the processing resources. According to the mentioned above results, these ratios are calculated and listed in Table \ref{tblResults}. For example in case of Scenario 1, for Linear topology this ratio is calculated by dividing the \textit{diff\_thpt} by the \textit{diff\_util} (shown in Fig. \ref{sim_scen1}).  
\begin{table}[!t]
	\renewcommand{\arraystretch}{1.3}
	\caption{Ratio of differences of throughputs to utilizations between our scheduler and default scheduler}
	\label{tblResults}
	\centering
	\begin{tabular}{c c c c}
		\hline
		\bfseries Scenario & \bfseries Linear & \bfseries Diamond & \bfseries Star \\
		\hline
		1 & 1.41 & 2.67 & 1.15 \\
		2 & 1.15 & 1.03 & 1.17 \\
		3 & 2.68 & 1.95 & 1.52 \\
		\hline
	\end{tabular}
\end{table}

\section{Related Works}
Many scheduling algorithms are proposed to optimize throughput or resource utilization in Storm. One of the primary considerations in task assignment is communication cost; Aniello et al. proposed two schedulers for Storm [3]. Both schedulers find hot edges of topology and put related tasks alongside each other, but the difference is on the way they determine hot edges. In R-Storm [6], B. Peng \textit{et al}. have presented a more intelligent scheduling algorithm. This algorithm is resource-aware and classifies resource constraints to soft and hard. R-Storm considers the CPU usage of homogenous machines and cannot be applied in heterogeneous environments, because it uses the same unit for CPU utilization of different machines. Rychl et al. [10] presented a heterogeneity-aware scheduler for the stream processing frameworks. However, the main idea of this scheduler is based on design-time knowledge as well as benchmarking techniques. As a drawback, this scheduler makes the scheduling decision by trial and error and gives no information about the optimum number of instances. 
\par
Some other papers deal with dynamic elasticity; chiefly [7], [9], [17], [18] change the scalability by tyning the parallelism degree of graph nodes. They monitor the data transfer rate of graph edges to determine its required parallelism. In these papers the total number of available machines is unlimited. B. Lohrmann \textit{et al.} [8] focused on latency constraints and presented a runtime elastic strategy that monitors and measures necessary metrics to make proper scale-up or scale-down operation. V. D. Veen \textit{et al.} [19] design and implement elasticity at the virtual machine level. They monitor several metrics such as processor load, the size of an input queue, and the number of tuples emitted and acknowledged to make the decision about whether to increase the number of virtual machines or decrease it. Stela [9] presents an on-demand mechanism to do scale-out or scale-in operation, depending on user request. However, in the Stela parallelism can be changed dynamically, but the total number of tasks is defined by the user. D-Storm [20] models the scheduling problem as a bin-packing problem and proposes a First Fit Decreasing (FFD) heuristic algorithm to solve it. As a goal, they try to minimize inter-node communication; by dynamic rescheduling and do not consider the overhead of reassignment of the tasks. I-Scheduler [26] is another scheduling algorithm which tries to overcome the complexity of the task-assignment problem by reducing the total number of tasks. It finds highly communicating tasks by exploiting graph partitioning techniques and take only one instance for each of them. In this way, it reduces the total number of tasks and overall communication cost. 
\par
In both scheduling and scaling solutions, a primary execution graph is submitted to Storm to get run; and the user determines the number of instances for each component of this graph. In the case of Storm, the number of instances is defined before running the topology, and we have to restart the entire system to change that. Some algorithms like Stela [9], take a specific number as the maximum number of instances and change the number of instances running on one executer, to scale up or down at run-time. Whereas, our algorithm offers a fitting execution graph for a cluster of heterogeneous machines. In the case of changing the number of machines, it can be re-executed to create new execution graph.

\section{Conclusions and Future Directions}
At the age of heterogeneous computing, it is extremely important to consider the different computing power of processors in a distributed environment. Apache Storm has a default scheduler which assigns tasks to processing elements regardless of their capabilities, therefore it results in under-utilization of processing resources in heterogeneous clusters. 
\par
To overcome this deficiency, we proposed a scheduling algorithm which takes the different computing power of heterogeneous machines into account. Our heterogeneity-aware algorithm, tries to maximize overall throughput, by generating a fitting execution topology graph for a given cluster. It also guarantees that no machine will be over-utilized when the topology is running. During the execution by any change in the cluster state, this algorithm can be used to recalculate the new number of instances and their suitable assignment. Nevertheless, some of Storm limitations prevent achieving the best throughput or appropriate resource utilization in heterogeneous systems. As our experiments show, one of the most significant obstacles to gain maximum CPU utilization is simple grouping strategies of Storm. 
\par
Our future work mainly aims at possible improvements of the scheduler efficiency, at addressing the obtaining profiling data problem, and at addressing the issues related to the load balancing. To handle the load balancing problem, we are implementing an intelligent grouping method which determines adequate rates for each task, depending on the computation power of the machine in which the task is running on.

\ifCLASSOPTIONcaptionsoff
  \newpage
\fi

\begin{IEEEbiography}[{\includegraphics[width=1in,height=1.25in,clip,keepaspectratio]{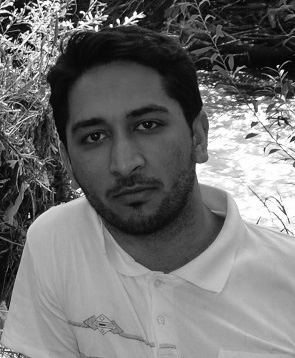}}]{Hamid Nasiri}
	is a Ph.D. candidate at the Department of Computer Engineering,	Sharif University of Technology, Tehran, Iran. He received the B.Sc. degree in Computer Engineering from Shahid Bahonar University of Kerman and the M.Sc. in Computer Architecture from Sharif University of Technology in 2013 and 2015, respectively. His research interests include big data stream processing, cloud computing and reconfigurable computing.
\end{IEEEbiography}

\begin{IEEEbiography}[{\includegraphics[width=1in,height=1.25in,clip,keepaspectratio]{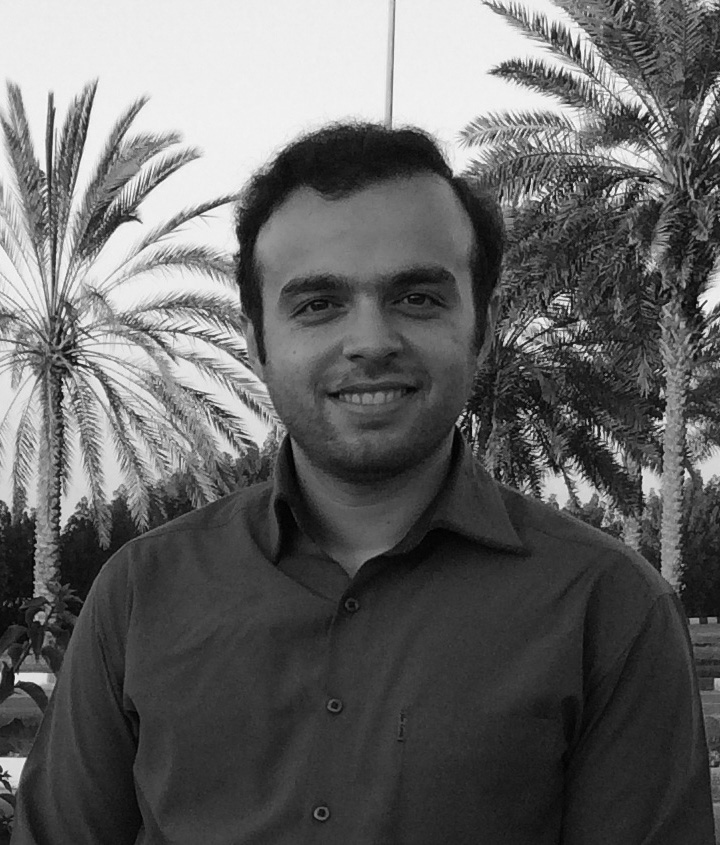}}]{Saeed Nasehi}
	received bachelor and master's degrees from University of Zanjan in 2015 and Sharif University of Technology in 2017, respectively. His current research focuses on the cloud computing, big data and distributed processing.
\end{IEEEbiography}

\begin{IEEEbiographynophoto}{Arman Divband}
	received bachelor and master's degrees from Iran University of Science and Technology in 2015 and Sharif University of Technology in 2018, respectively. His research interests include big data processing and green computing.
\end{IEEEbiographynophoto}  

\begin{IEEEbiography}[{\includegraphics[width=1in,height=1.25in,clip,keepaspectratio]{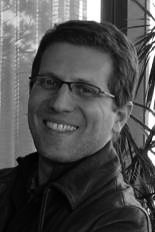}}]{Maziar Goudarzi}
	(S’00, M’11, SM'16) is an Associate Professor at the Department of Computer Engineering, Sharif University of Technology, Tehran, Iran. He received the B.Sc., M.Sc., and Ph.D. degrees in Computer Engineering from Sharif University of Technology in 1996, 1998, and 2005, respectively. His current research interests include architectures for large-scale computing systems, green computing, hardware-software co-design, and reconfigurable computing. 
\end{IEEEbiography}

\end{document}